\pdfoutput=1
\documentclass[usenatbib,usegraphicx]{mn2e}
\usepackage[breaklinks,colorlinks,citecolor=blue,linkcolor=magenta]{hyperref}
\usepackage{amssymb}
\usepackage{graphicx}
\usepackage{epsfig}
\usepackage{color}
\usepackage{longtable}

\newcommand{\OVIdblt}{{O}\kern 0.1em{\sc vi}~$\lambda\lambda 1032, 1038$}
\newcommand{\CII}{\hbox{{C}\kern 0.1em{\sc ii}}}
\newcommand{\CIII}{\hbox{{C}\kern 0.1em{\sc iii}}}
\newcommand{\CIV}{\hbox{{C}\kern 0.1em{\sc iv}}}
\newcommand{\HI}{\hbox{{H}\kern 0.1em{\sc i}}}
\newcommand{\Lya}{\hbox{{Ly}\kern 0.1em$\alpha$}}
\newcommand{\Lyb}{\hbox{{Ly}\kern 0.1em$\beta$}}
\newcommand{\Lyg}{\hbox{{Ly}\kern 0.1em$\gamma$}}
\newcommand{\Lyd}{\hbox{{Ly}\kern 0.1em$\delta$}}
\newcommand{\Lye}{\hbox{{Ly}\kern 0.1em$\epsilon$}}
\newcommand{\Lyz}{\hbox{{Ly}\kern 0.1em$\zeta$}}
\newcommand{\Lyeta}{\hbox{{Ly}\kern 0.1em$\eta$}}
\newcommand{\MgII}{\hbox{{Mg}\kern 0.1em{\sc ii}}}
\newcommand{\OVI}{\hbox{{O}\kern 0.1em{\sc vi}}}
\newcommand{\OVII}{\hbox{{O}\kern 0.1em{\sc vii}}}
\newcommand{\OVIII}{\hbox{{O}\kern 0.1em{\sc viii}}}
\newcommand{\NV}{\hbox{{N}\kern 0.1em{\sc v}}}
\newcommand{\SiII}{\hbox{{Si}\kern 0.1em{\sc ii}}}
\newcommand{\SiIII}{\hbox{{Si}\kern 0.1em{\sc iii}}}
\newcommand{\SiIV}{\hbox{{Si}\kern 0.1em{\sc iv}}}
\newcommand{\FeII}{\hbox{{Fe}\kern 0.1em{\sc ii}}}
\newcommand{\kms}{\ensuremath{\mathrm{km~s^{-1}}}}

\newcommand{\NeIX}{\hbox{{Ne}\kern 0.1em{\sc ix}}}
\newcommand{\OI}{\hbox{{O}\kern 0.1em{\sc i}}}

\def\apj{ApJ}

\def\aap{A\&A}

\def\mnras{MNRAS}

\def\pasj{PASJ}

\title[SNRs in Turbulent MCs]
{Numerical Simulations of Supernova Remnants in Turbulent Molecular Clouds}

\author[Zhang \& Chevalier]
{Dong Zhang$^{1}$\thanks{E-mail: dz7g@virginia.edu} and Roger A.~Chevalier$^{1}$\\
$^1$Department of Astronomy, University of Virginia, 530 McCormick Road, Charlottesville, VA 22904, USA}

\begin{document}

\maketitle

\begin{abstract}

Core-collapse supernova (SN) explosions may occur in the highly inhomogeneous molecular clouds (MCs) in which their progenitors were born. We perform a series of 3-dimensional hydrodynamic simulations to model the interaction between an individual supernova remnant (SNR) and a turbulent MC medium, in order to investigate possible observational evidence for the turbulent structure of  MCs. We find that the properties of SNRs are mainly controlled by the mean density of the surrounding medium, while a SNR in a more turbulent medium with higher supersonic turbulent Mach number shows lower interior temperature, lower radial momentum, and dimmer X-ray emission compared to one in a less turbulent medium with the same mean density. We compare our simulations to observed SNRs, in particular, to W44, W28 and IC 443. We estimate that the mean density of the ambient medium is $\sim 10\,$cm$^{-3}$ for W44 and W28. The MC in front of IC 443 has a density of $\sim 100\,$cm$^{-3}$. We also predict that the ambient MC of W44 is more turbulent than that of W28 and IC 443. The ambient medium of W44 and W28 has significantly lower average density than that of the host giant MC. This result may be related to the stellar feedback from the SNRs' progenitors. Alternatively, SNe may occur close to the interface between molecular gas and lower density atomic gas. The region of shocked MC is then relatively small and the breakout into the low density atomic gas comprises most of the SNR volume.


\end{abstract}


\section{Introduction}

An important class of core-collapse supernova remnants (SNRs) has been found to be interacting with molecular clouds (MCs), presumably the birth clouds of the supernova progenitors. Direct evidence for SNR-MC interaction comes from observations of the OH (1720 MHz) maser transition line, which is a powerful tracer associated with SNR shock waves in MCs (\citealt{frail94,claussen97,frail98,claussen99}; \citealt{Wardle02} and references therein). More generally, SNR-MC interaction can be shown by broad molecular line features, including H$_2$, CO, HCO$^+$, HCN, and CS (\citealt{Seta98,Seta04,reach05,Jiang10,Lee12}; \citealt{slane15} and references therein).

There is a strong correlation between SNRs that show molecular emission and those that have a mixed morphology. The term ``mixed morphology" (also called ``thermal composite") is used for SNRs that show a shell structure at radio wavelengths, but center-filled emission at X-ray wavelengths (\citealt{rho98,Jones98}). Among the known 295 Galactic SNRs (\citealt{Green14}), about 40 SNRs  belong to this category (\citealt{Lazendic06,Vink12}).
Good examples of this type of remnant are W44, W28 and IC 443, all of which are known to be interacting with MCs. Several generations of X-ray satellites have been used to observe the detailed X-ray properties of these SNRs (see Section \ref{section_observations} for references). The X-ray emission is primarily thermal, so it gives information on the hot gas content of the remnants.
A more recent finding is the evidence for a recombining X-ray spectrum, which has been observed for W44 \citep{uchida12}, W28 \citep{sawada12}, and IC 443 \citep{Yamaguchi09,Ohnishi14} among others.
Furthermore, many of these remnants have been detected as sources of high energy $\gamma$-ray emission, both at GeV ({\it Fermi} and {\it AGILE}) and TeV (ground-based Cherenkov detectors) energies (e.g., \citealt{Abdo10a,Abdo10b,Abdo10c,Tang14,Humensky15}).

The origin of the mixed morphology SNRs is still unclear. A variety of analytic models beyond the standard SNR model (e.g., \citealt{Sedov59,chevalier77,Draine11}) have been proposed for these SNRs. These analytic models describe dense clumps surrounded by the interclump ambient medium with a low density. \cite{white91} developed self-similar solutions for the interaction of a blast wave with  clumps, showing that evaporation of the clumps by thermal conduction could lead to center filled thermal X-ray emission. In this model, the blast wave moves rapidly through the interclump medium and does not cool.
\cite{chevalier99} proposed a model for SNRs primarily propagating in the interclump medium with a density $\sim 5-25$\,cm$^{-3}$ and interacting with molecular clumps with a density $\sim 10^3$\,cm$^{-3}$. 
Models for W44 and IC 443 involved the supernova shock wave becoming radiative in the interclump region, and the interaction
of the cool shell with clumps giving rise to the molecular emission.
\cite{cox99} and \cite{shelton99} developed a partially radiative remnant model for W44, with an emphasis on the X-ray emission.
In this model, there is a gradient in the ambient density, which is $6$\,cm$^{-3}$ on average.
A key feature of the model is the action of thermal conduction in the hot gas.
\cite{reach05} presented molecular line observations of the remnants W44 and W28, leading to a model for W44 in which a non-radiative shock wave with velocity $500~\kms$ moves through an interclump medium with a density of $5$ cm$^{-3}$.


An important feature of MCs is turbulence. 
Molecular lines from MCs show a correlation between the length of MCs ($L$) and the widths of lines ($\sigma$) that $\sigma \propto L^{\epsilon}$, with $\epsilon\sim 0.2-0.5$ (\citealt{Larson81,Heyer09,Lombardi10,Schneider11}). The $\sigma-L$ relation shows that molecular lines have features that indicate the presence of supersonic turbulence (\citealt{Zuckerman74,Falgarone90}).
Such turbulent motions can generate a turbulent density structure,
 and there is evidence for
such structure from the column density distribution of CO and dust in MCs \citep{padoan97} and the
density of H$_2$CO \citep{ginsburg13}. On the other hand, numerical simulations have been performed to study turbulence structure for two decades (\citealt{Gammie96}). Simulations show that for isothermal gas, the density probability distribution function (PDF) with respect to mass or volume  tends to a lognormal distribution in the case of supersonic turbulence \citep{vazquez94,Ostriker01,kevlahan09,lemaster09a,lemaster09b}.
The peak of the distribution with respect to mass gives the density of clumps that have most of the mass, while the peak of the distribution with respect to volume gives the density that occupies most of the volume.

Recently, a series of 3-dimensional (3D) numerical simulations have been carried out to investigate the evolution of SNRs interacting with a multiphase or turbulent medium.
\cite{martizzi15} considered SNR expansion into a density structure with the lognormal density distribution
expected from supersonic turbulence, quantified the momentum and energy injection from individual SNRs into the turbulent ISM, and developed a series of analytic formulae that provide the ``sub-grid" models for galaxy-scaled simulations of SN-driven outflows \citep{martizzi16,Fielding17}.
\cite{walch15}  considered a fractal density structure and a lognormal probability distribution function (PDF), and studied the SNR-MC interaction.
\cite{iffrig15} allowed a presupernova turbulent velocity field to act, leading to
the assumed density structure, with a declining PDF at high density more like a power
law than a lognormal distribution.
\cite{kim15} also simulated 3D remnants, but in a 2-phase interstellar medium, not
a turbulent density structure. All the above works had the aim of determining the supernova remnant momentum
and energy feedback in the ISM and MCs for large scale simulations of galaxy formation and evolution.
The emphasis was thus on the late times, with little attention to the appearance of middle-aged remnants.

Other numerical simulations directly modeled synthetic emission from remnants and compared to observations.  \cite{Slavin17} presented 2D and 3D hydrodynamic simulations including thermal conduction to test the analytic model of \cite{white91}. The effects of non-thermal broadband emission were simulated by \cite{Ferrand14}. 
Simulations have been directly compared to observations of SN 1987A  (\citealt{Orlando15}), Cassiopeia A (\citealt{Orlando16}), and CTB 109 \citep{Bolte15}.
However, turbulent molecular clouds were not involved in these works. 


Our purpose in this paper is to perform hydrodynamic simulations to model SNR interacting with a turbulent medium, in order to see whether the interaction shows observational evidence for the turbulent structure of the surrounding clouds. We organize the paper as follows. We discuss the models of MC turbulence in Section \ref{section_turb},  radiative cooling and X-ray broadband emission in Section \ref{section_cooling}, and the simulation setup for our work in Section \ref{section_setup}. The simulation results are shown in Section \ref{section_hydroresults}, and synthetic X-ray images and light curves are shown in Section \ref{section_Xray}. In Section \ref{section_observations} we compare our models to observations of remnants, in particular, to W44, W28 and IC 443. Conclusions and discussion are  in Section \ref{section_conclusions}.

\section{Numerical Method and Simulation Setup}

\subsection{Turbulent Molecular Clouds}\label{section_turb}

\begin{figure}
\centerline{\includegraphics[width=8.5cm]{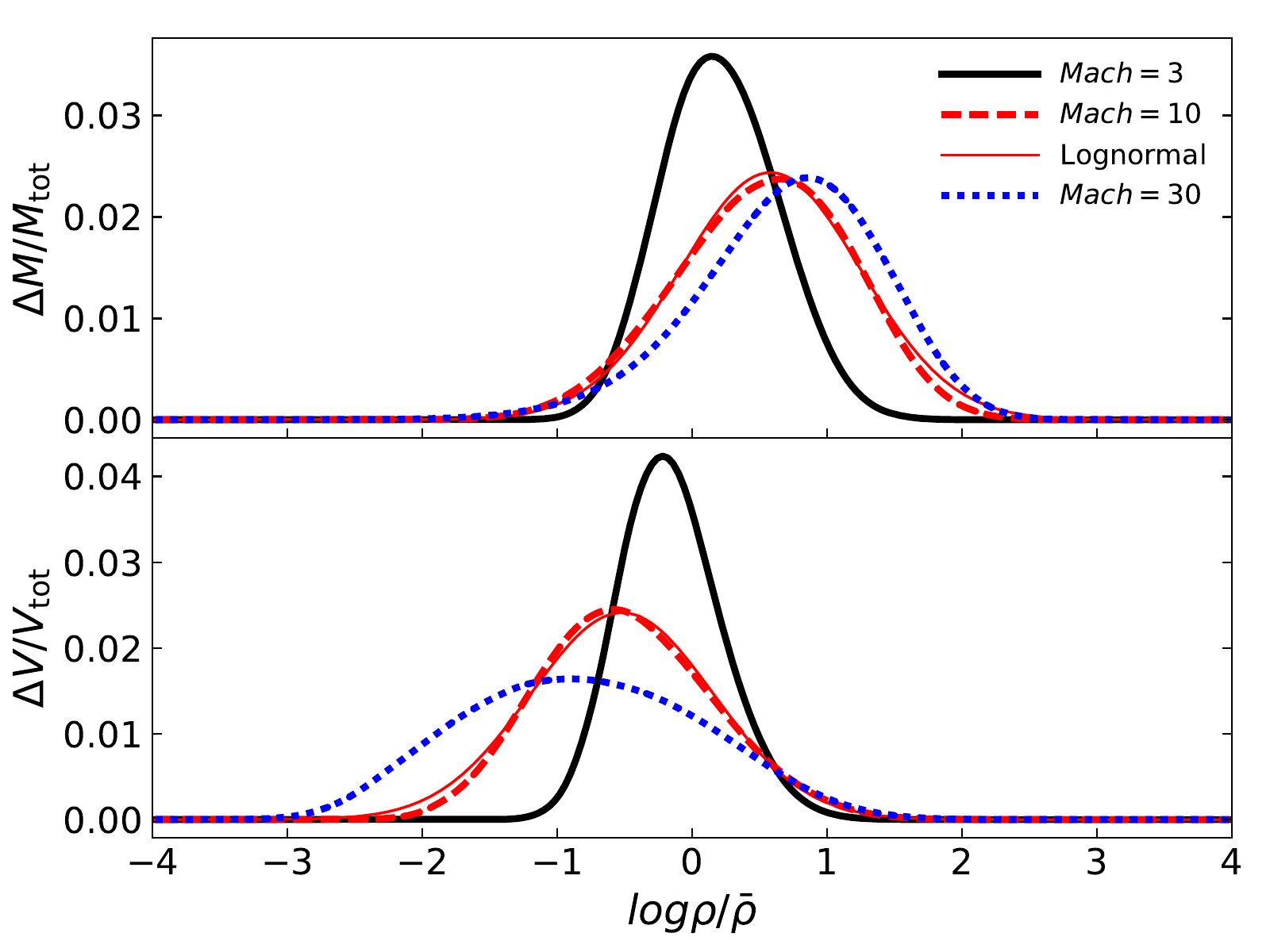}}
\caption{Density (upper) and volume (lower) probability distribution functions of MCs with initial supersonic turbulent Mach number $\mathcal{M}=3$ (solid lines), 10 (dashed lines) and 30 (dotted lines). The lognormal distributions are shown to compare to the numerical results for the $\mathcal{M}=10$ run, where we choose $\mu_{\rm M}=0.57$ for density and $\mu_{\rm V}=-0.52$ for volume.}\label{fig_PDF}
\end{figure}

Giant molecular clouds (GMCs) are highly inhomogeneous (\citealt{Hennebelle12} and references therein). It is still unknown what mechanisms drive supersonic turbulence in GMCs. Gravity, star formation feedback, and magnetic fields may play a significant role in maintaining the turbulence, which is expected to rapidly dissipate without energy injection (\citealt{MacLow04,KB16}). In this paper we neglect the energy source of turbulence, and use a ``decaying model," in which the turbulence is only driven at time $t=0$ and then decays without further turbulence driving, to describe the turbulent structure in GMCs. Following \cite{Ostriker01} (see also \citealt{Stone98}; \citealt{lemaster09a, lemaster09b}), we perform a series of 3D turbulence simulations with an initially uniform, stationary gas and assume an isothermal equation of state for the gas. The initial velocity perturbation is generated at $t=0$ following a Gaussian random distribution with a Fourier power spectrum $|v^{2}(k)|\propto  k^{-4}$ with $k$ being the wavenumber between two cutoffs $k_{\rm min}$ and $k_{\rm max}$. We set $k_{\rm min}L/(2 \pi)=1$ and $k_{\rm max}L/(2 \pi)=N/2$, where $L$ is the length of the box, and $N$ is the number of zones in each direction. We use $N=512$ for most of the simulations. The turbulence is characterized by the turbulent Mach number ${\mathcal{M}}=\langle \sigma_v^{2}/c_{s}^{2}\rangle^{1/2}$, where $c_s$ is the sound speed and $\sigma_v$ is the velocity dispersion in the gas.  We use the turbulence driver implemented in the codes \textsc{athena} and \textsc{athena++}  to carry out simulations with initial supersonic Mach number $\mathcal{M}=3,10$ and 30, and let the turbulence decay with time, creating inhomogeneous GMCs. Figure \ref{fig_PDF} shows the snapshots of the density probability distribution function (PDF) at $t/t_*=50$, where $t_*=L/(N\mathcal{M}c_s) = t_s/(N\mathcal{M})$ is the time it takes the flow to cross one zone, and $t_s=L/c_s$ is the sound crossing time for the computational box. Much of the mass is in the high density gas and much of the volume is in the low density gas. The density and volume PDFs of the turbulent MCs can be approximately fitted with lognormal distributions:  
\begin{equation}
f_{V,M}(y)dy=\frac{1}{\sqrt{2\pi \sigma^2}}\exp\left[\frac{-(y\pm\mu)^2}{2\sigma^2}\right]dy,\label{lognormal}
\end{equation}
where $y=  \log \rho /\bar\rho$ and $\bar\rho$ is the mean density.  The mean and dispersion $\mu$ and $\sigma$ are related by $\sigma =\sqrt{2\mu/\ln 10} \approx 0.93\sqrt{|\mu|}$ (\citealt{Ostriker01}). We find that  for density PDFs $\mu_{\rm M}\approx 0.25$ and $\sigma_{\rm M}\approx 0.47$ for $\mathcal{M}=3$, and $\mu_{\rm M}\approx 0.57$ and $\sigma_{\rm M}\approx 0.70$ for $\mathcal{M}=10$. Turbulence with higher Mach number decays faster than that with lower Mach number, so the density and volume profiles of the run with $\mathcal{M}=30$ are only slightly different from the run with $\mathcal{M}=10$.  Figure \ref{fig_PDF} also shows a comparison of the lognormal distributions with the numerical results for the $\mathcal{M}=10$ run, where we use $\mu_{\rm M}=0.57$ and $\mu_{\rm V}=-0.52$.
In addition to the decaying model for turbulence simulations, the ``energy injection model" with continuous energy injection into the turbulent medium (\citealt{lemaster09a}) will be discussed in Section \ref{section_observations}.


\begin{figure}
\centerline{\includegraphics[width=8.5cm]{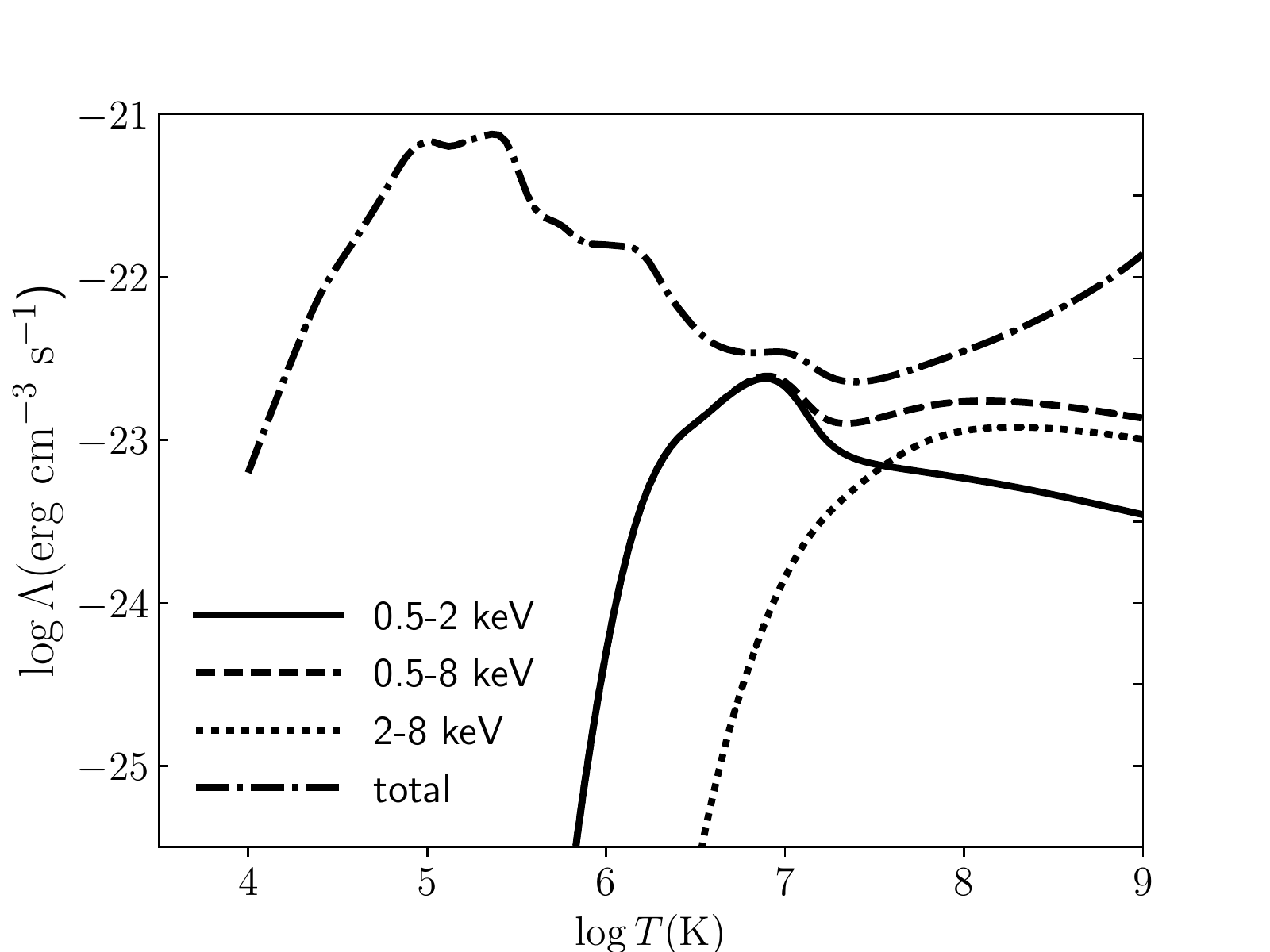}}
\caption{Total cooling function calculated by \textsc{cloudy}, and broadband cooling functions calculated by \textsc{spex} in different frequency ranges for solar metallicity. }\label{fig_cooling}
\end{figure}

\subsection{Radiative Cooling and X-ray Emission}\label{section_cooling}

For the interaction between a SNR and a turbulent MC, a crucial part is the allowance for radiative cooling of the shocked gas. The 1D spherical case of a cooling SNR in a uniform medium has been well studied \citep{chevalier74,cioffi88,Draine11}. A basic result is the expansion of a SNR occurs several stages. A SNR first expands freely until its swept-up mass becomes comparable to its initial ejected mass. Then, the SNR moves to the blast wave stage -- the so-called Sedov-Taylor (ST) stage with a self-similar structure. After radiative cooling becomes important the SNR goes to the third and longest stage -- the radiative cooling stage or the ``snow-plow" stage. The radiative cooling stage can be divided into the early pressure-driven snow plow stage in which the thermal pressure of the SNR cannot be neglected, and the late momentum-conserving stage.  Eventually the SNR forward front shock velocity decreases to the sound speed of the ambient medium.  The multiple stages of SNR evolution have been confirmed by recent 3D numerical simulations (\citealt{kim15,walch15}). Following these numerical simulations, we model radiative cooling of SNR in two different gas temperature regimes. For gas temperature $T\geqslant 10^{4}\,$K, we use the \textsc{cloudy} code version 17.00\footnote{http://www.nublado.org/} (\citealt{Ferland17}) to generate the collisional ionization equilibrium (CIE) cooling function $\Lambda$ (in  units of erg\,cm$^3$\,s$^{-1}$) for solar metalicity. The treatment of CIE is simplified, as the radiative interface may not be in ionization equilibrium. Treatments of non-equilibrium ionization (NEI, \citealt{Kafatos73,Shapiro76,MacDonald81,Sutherland93,Harrus97}) have been used to give the detailed interface structure. However, it was shown that these details are irrelevant to the global evolution of a remnant, and CIE is a good approximation to model the dynamics of the remnant (\citealt{Bertschinger86,cioffi88}). 
For low gas temperature $T < 10^{4}\,$K, we adopt the analytic cooling function given by \cite{Koyama02} for simplicity:
\begin{eqnarray} \label{EQ_KI}
 \Lambda & =& \Gamma  \left[10^7 {\rm exp}\left(\frac{-1.184 \times 10^5}{T + 1000} \right) \right. \nonumber \\
& &+ \left. 1.4 \times 10^{-2} \sqrt{T}\; {\rm exp}\left(\frac{-92}{T} \right)\right] \frac{\rm erg\; cm^3}{\rm s} \label{EQ_KI}
\end{eqnarray}
where $\Gamma$ is the heating rate $\Gamma = 2\times 10^{-26}$ ergs\,s$^{-1}$. Equation (\ref{EQ_KI}) was derived from analytic fitting of the Lyman $\alpha$ and C$^{+}$ cooling rates at solar metallicity.  Our results are not sensitive to the details.


In order to compute X-ray emission from SNRs (see Section \ref{section_Xray}), we also need to obtain cooling functions in the X-ray bands. Here, we use the  \textsc{spex} package\footnote{http://www.sron.nl/spex}  (\citealt{Schure09}) to generate the broadband emission between $0.5-2\,$keV (soft X-ray), $2-8\,$keV (hard X-ray) and $0.5-8\,$keV (see also \citealt{Zhang14}).  Note that the X-ray emission from SNRs has long been modeled as  NEI emission. As discussed by \cite{Schure09}, the NEI cooling rates are lower than that of a CIE cooling curve, but the differences between NEI and CIE emission are by a factor of $\lesssim 1.5$ for $T\geq 10^6\,$K. We  use the CIE treatment as an approximation.  Figure \ref{fig_cooling} shows the total cooling function $\Lambda$ and broadband emission as functions of temperature.

\subsection{Simulation Setup}\label{section_setup}

We perform a series of simulations in a 3D computational box using the \textsc{athena++} code, which used the Harten-Lax-van Leer-Contact (HLLC) Riemann solver and a second-order van Leer integrator (VL2) for hydrodynamics (\citealt{White16}). The ambient turbulent MC as the background in the box is generated by the turbulence driver as introduced in Section \ref{section_turb}. Table \ref{tab_parameters} summarizes parameters of the turbulent MC background in this paper. We chose $\bar{n}_{\rm H}=100\,$ cm$^{-3}$ and the gas temperature of $T=30\,$K as the typical mean density for MCs (\citealt{Blitz93,Williams00}), with $\mathcal{M}=10$ as the fiducial run (MC1).  The other two turbulent MC models have $\mathcal{M}=3$ (MC2) and $\mathcal{M}=30$ (MC3), as given in Section \ref{section_turb}. The box size is $L=16\,$pc for $\bar{n}_{\rm H}=100\,$cm$^{-3}$. We also performed two other sets of simulations: one with the ambient mean density $\bar{n}_{\rm H}=10\,$cm$^{-3}$, $T=300\,$K with a box size of $L=64\,$pc, and another with the interstellar medium mean density $\bar{n}_{\rm H}=1\,$cm$^{-3}$, $T_{\rm ISM}=2500\,$K and a box size of $L=128\,$pc. In all cases, the ambient temperature and pressure were sufficiently low that they do not affect the dynamics of a middle-aged SNR. The  zone size is $\Delta x = L/N$, with a fiducial zone number of $N=512$.  Therefore $\Delta x = 1/32\,$pc, 1/8\,pc and 1/4\,pc for $\bar{n}_{\rm H}=100\,$cm$^{-3}$, 10\,cm$^{-3}$ and 1\,cm$^{-3}$ respectively.  An individual SNR is located at the center of the box with an initial radius of $20\Delta x$, a total thermal energy of $10^{51}\,$ergs, and an initial mass of $3\,M_{\odot}$. The initial energy and size of the SNR are similar to those in \cite{kim15}. We reset the start time to $t=0$ as the SNR begins to expand.

The most straightforward test of the SNR evolution is the adiabatic Sedov-Taylor (ST) solution for a spherical blast wave in a uniform medium. Therefore, we also carry out a ``uniform run" to test the analytic ST model. Although the initial structure of the SNR with a radius of $20\Delta x$  is different from the ST solution, the SNR quickly converges to the ST solution. 
However, for a cooling SNR  even expansion in a uniform medium can be complex due to hydrodynamic instabilities  near the shock front \citep{sutherland03}. Our simulations with a resolution of $\Delta x$ from 1/32\,pc to 1/4\,pc do not resolve the cooling region and so are not accurate on these small scales.



The main feature of our simulations is the inclusion of a turbulent density structure and radiative cooling. Compared to the fiducial run, we also carry out an ``adiabatic run" with the turbulent structure and no cooling to separate out this particular aspect. This case would give an upper bound of the SNR size and swept-up mass compared to the runs with radiative cooling. 


Since the simulation resolution may not be high enough to resolve the structure of the interface,  a zone across interfaces between hot and cool gas is treated with an intermediate temperature gas temperature in the entire zone. Thus the emissivity in this zone is higher than the realistic case, in which only part of a zone has radiative cooling. There is thus the potential for overestimating the effects of radiative cooling in our simulations. We turn to convergence testing to check whether the simulation results depend on the resolution or not. We also carry out a low-resolution run with 256$^3$ zones and high-resolution run with 1024$^3$ zones to compare to the fiducial run. 

\begin{table}
\caption{Summary of Simulation Parameters}
\begin{tabular}{lccccccclcc}
\hline

Run & $\bar{n}_{\rm H}$ (cm$^{-3}$) & $\mathcal{M}$ & $L$ (pc) & $N^{3}$ \\
  \hline 
MC1      & 100 & 10 & 16 pc & 512$^3$ \\
MC2      & 100  & 3 & 16 pc & 512$^3$ \\
MC3      & 100  & 30 & 16 pc & 512$^3$ \\
$\bar{n}10$   & 10  & 10 & 64 pc  & 512$^3$\\
$\bar{n}1$    & 1  & 10 & 128 pc  & 512$^3$ \\
uniform      & 100  & 10 &  16 pc & 512$^3$ \\
adiabatic    & 100  & 10 & 16 pc & 512$^3$ \\
HR    & 100  & 10 & 16 pc  & 1024$^3$ \\
Low    & 100  & 10 & 16 pc  & 256$^3$ \\
\hline

\end{tabular}  \\
\noindent{\it Notes.}  The names MC$i$ stand for molecular cloud model $i$ with different initial inhomogeneous setups, and $\mathcal{M}$ is the turbulent  Mach number for the turbulent MC models. Runs $\bar{n}X$ have similar inhomogeneous setups to MC1 but the mean density of the molecular cloud background is $X\,$ cm$^{-3}$ and with different size of the 3D computational box.  Here $L$ is the size of the box, and $N$ is the number of zones in each direction of the box. 
\label{tab_parameters}
\end{table}


\section{Hydrodynamic Results}\label{section_hydroresults}

\begin{figure*}
\centerline{\includegraphics[width=18cm]{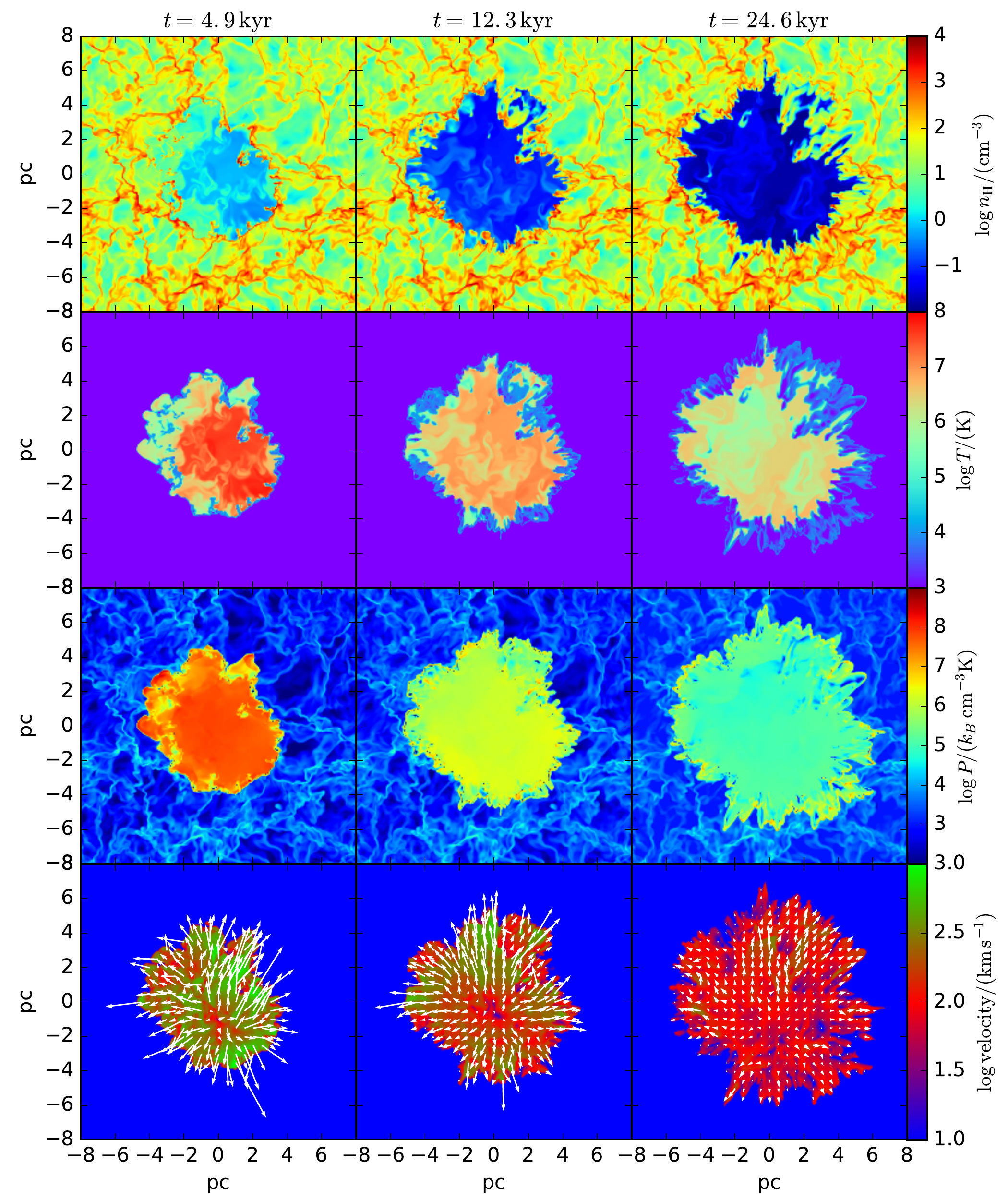}}
\caption{2D slices ($z=0$) of density, temperature, pressure and velocity at three snapshots $t\,=4.9\,$kyr, 9.8\,kyr and 18.4\,kyr for the fiducial run MC1. Note that the velocity fields in the velocity snapshots are the projected velocity on the plane of $z=0$.}\label{fig_turb100}
\end{figure*}
\begin{figure*}
\centerline{\includegraphics[width=18cm]{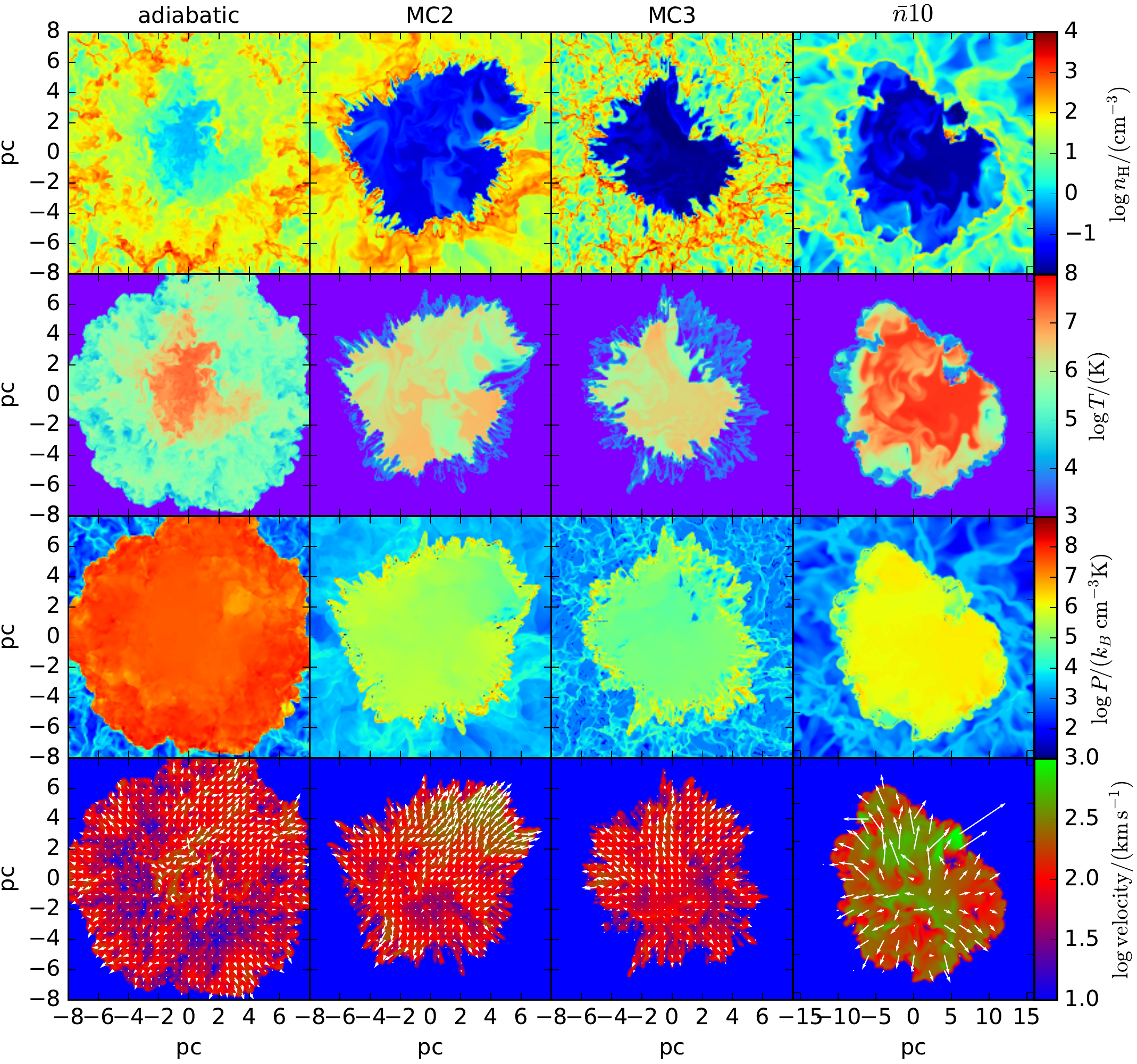}}
\caption{2D slices ($z=0$) of density, temperature, pressure and velocity for the adiabatic, MC2, MC3 and $\bar{n}_{10}$ runs. The definitions and parameters are given in Table \ref{tab_parameters}. Snapshots are taken at $t=12.3\,$kyr.}\label{fig_mixed}
\end{figure*}
\begin{figure*}
\centerline{\includegraphics[width=15cm]{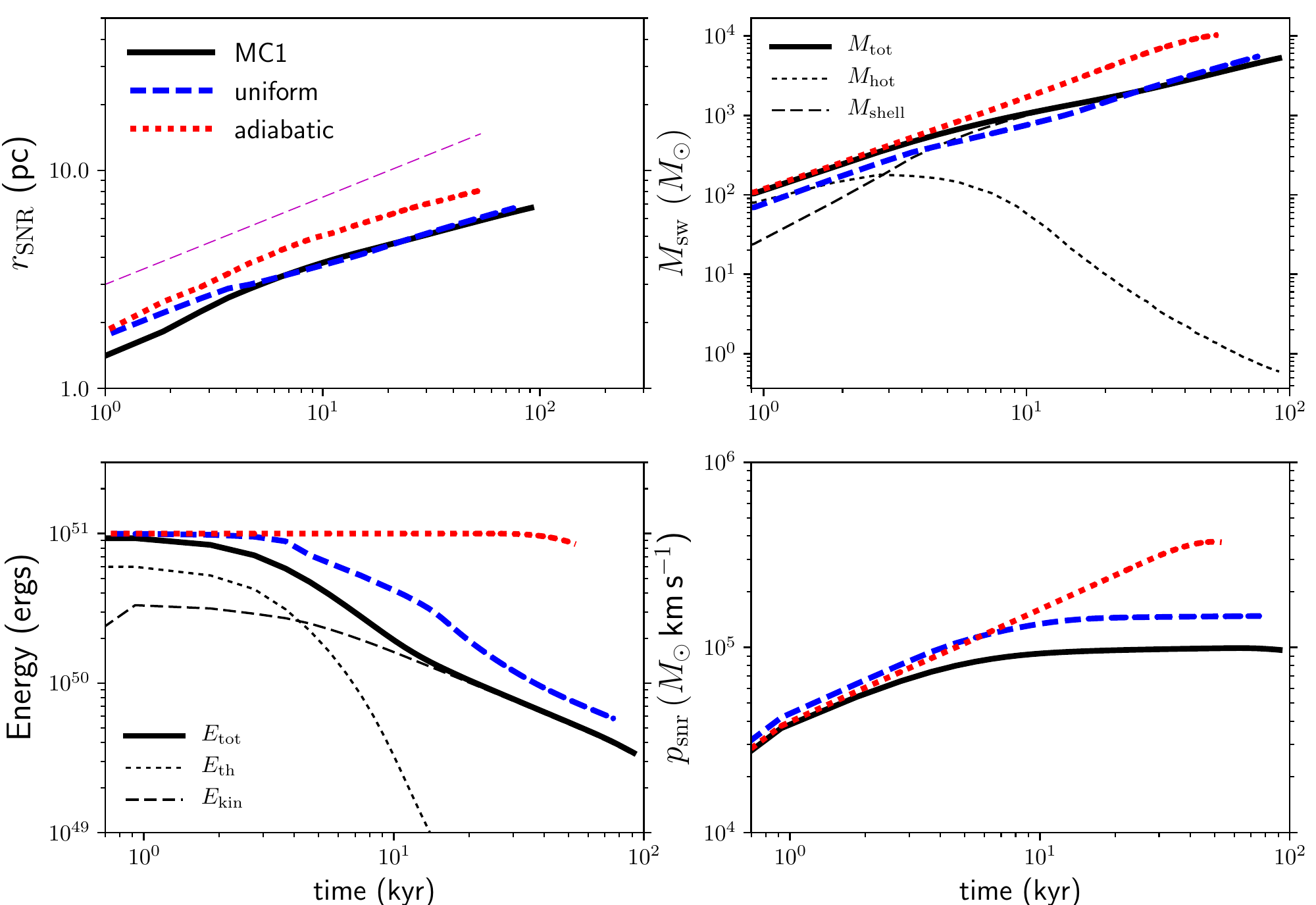}}
\caption{Time evolution of the averaged SNR shock front radii $r_{\rm SNR}$ (upper left), total swept MC mass (upper right),  total energy $E_{\rm tot}$ (lower left), and total radial momentum $p_{\rm SNR}$ (lower right) for MC1 (fiducial,black solid lines), the uniform (blue dashed lines) and the adiabatic (red dotted lines) runs. Note that for the fiducial run we also show the $r_{\rm SNR}\propto t^{0.4}$ relation in the upper left panel,  the mass of hot gas and shell $M_{\rm shell}$ in the upper right panel, and the thermal ($E_{\rm th}$) and kinetic ($E_{\rm k}$) energies in the lower left panel.}\label{fig_dyn1}
\end{figure*}
\begin{figure*}
\centerline{\includegraphics[width=15cm]{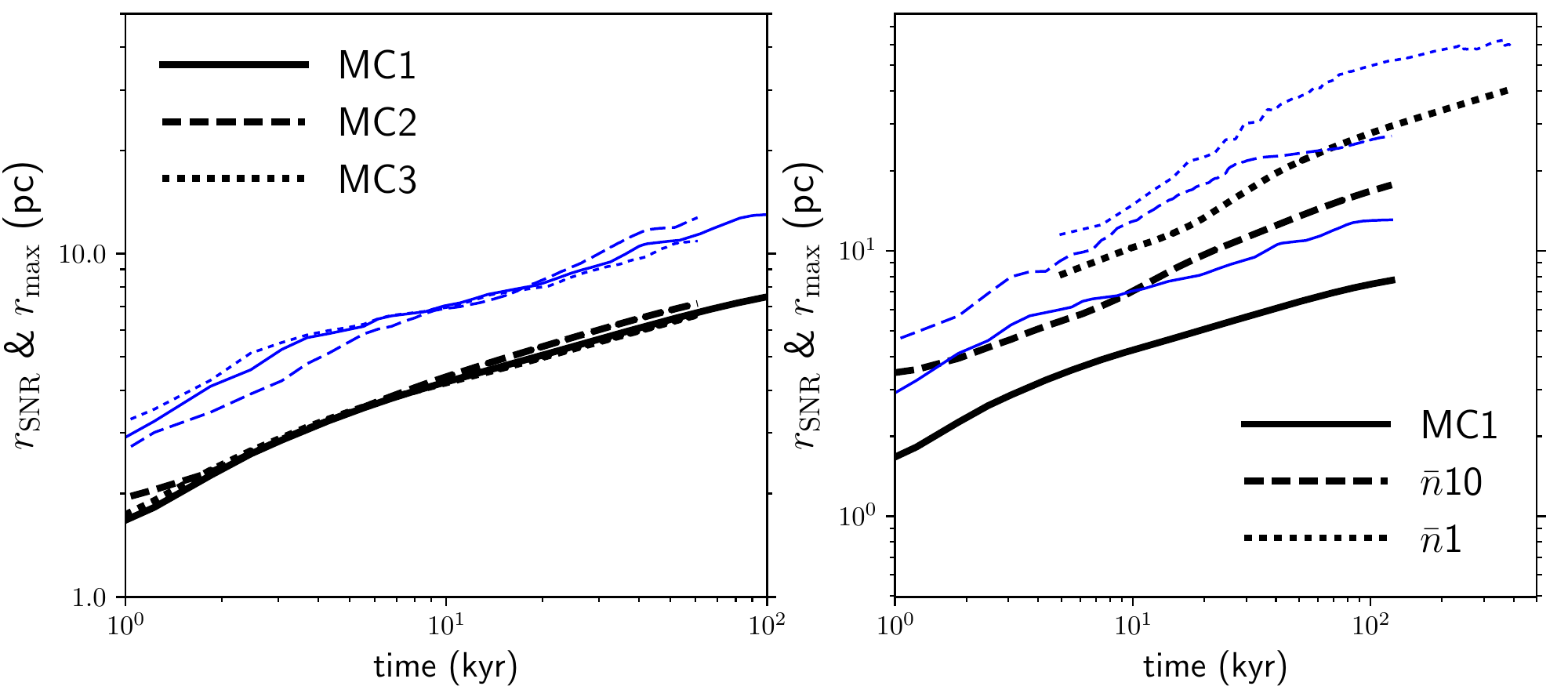}}
\caption{Comparison of the averaged SNR shock front radii $r_{\rm SNR}$ (black solid lines) and the maximum radii in the shell region $r_{\rm max}$ (blue thin lines) for the models MC1 (fiducial), MC2, and MC3 (left), and MC1, $\bar{n}10$ and $\bar{n}1$ (right).}\label{fig_Martizzi}
\end{figure*}
\begin{figure*}
\centerline{\includegraphics[width=15cm]{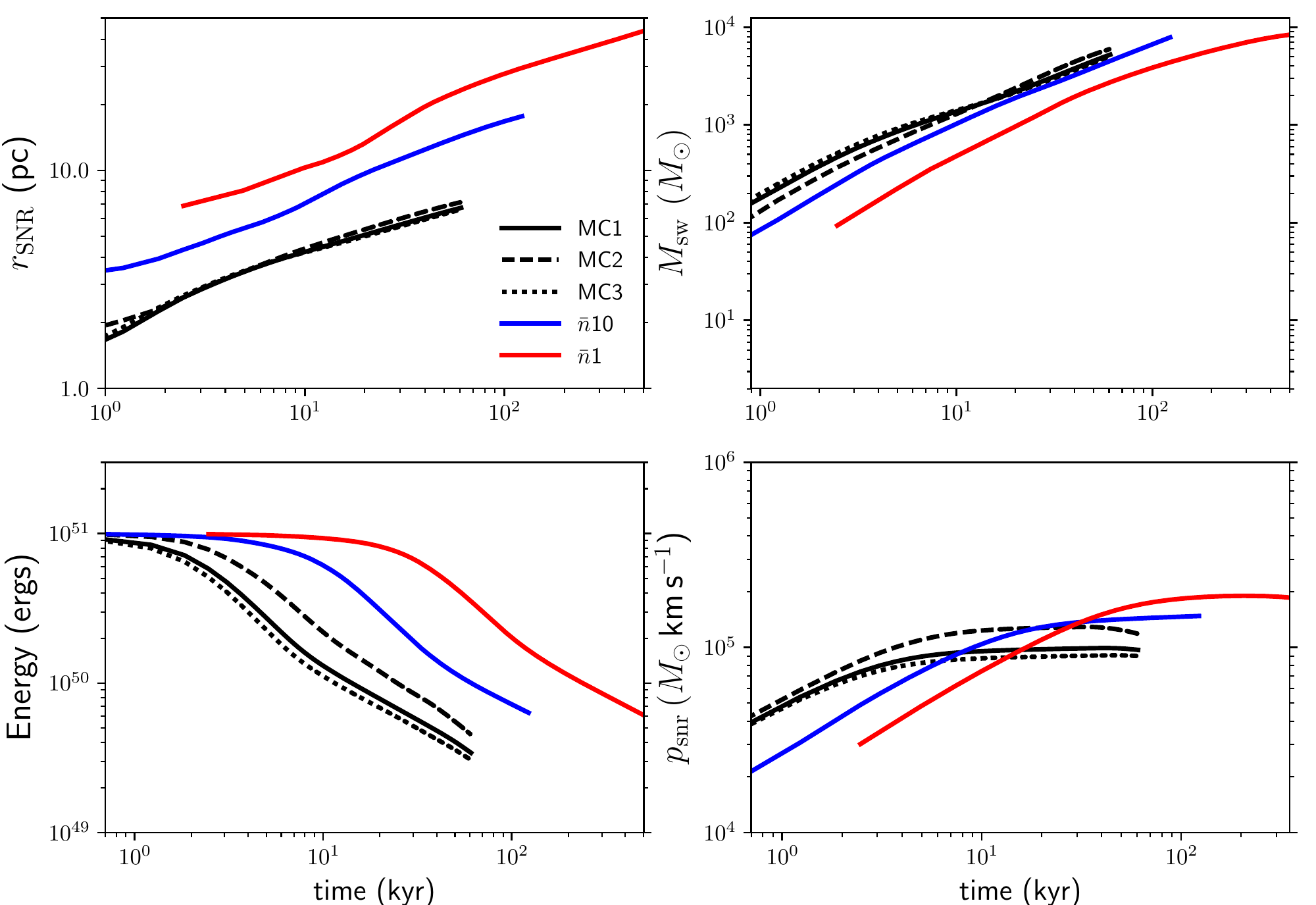}}
\caption{Similar as Figure \ref{fig_dyn1} but for MC1 (fiducial), MC2,  MC3, $\bar{n}$10 and $\bar{n}$1 runs.}\label{fig_dyn2}
\end{figure*}
\begin{figure*}
\centerline{\includegraphics[width=18cm]{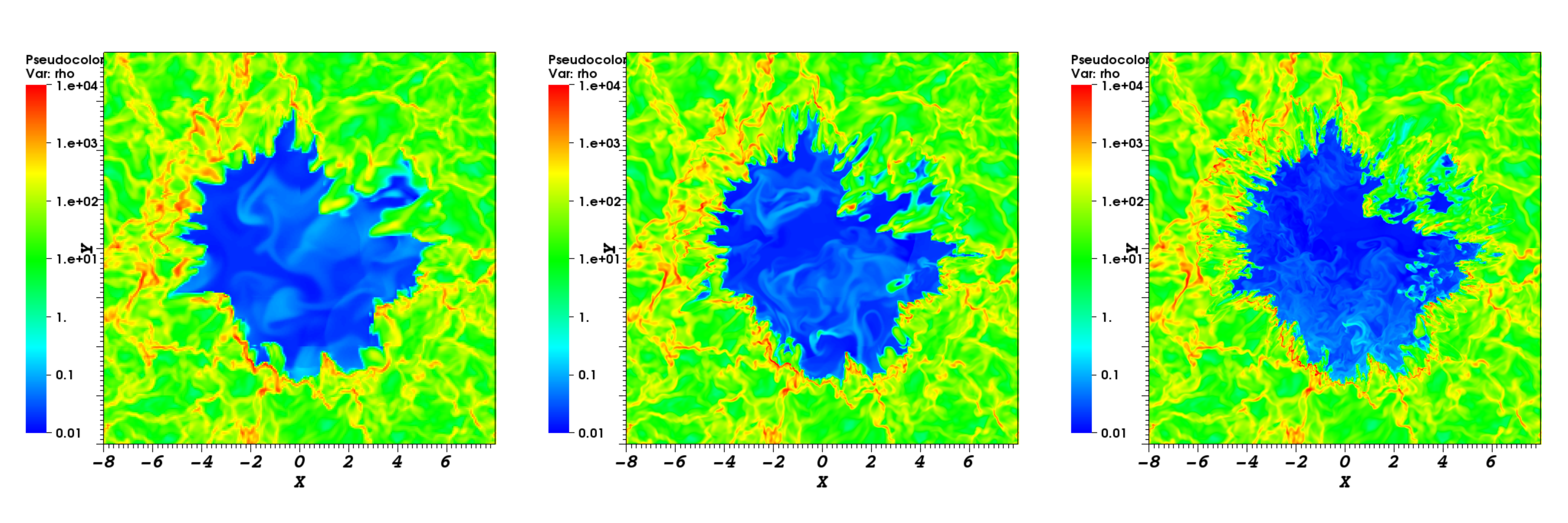}}
\caption{Comparison of density distributions (in unit of cm$^{-3}$) from the low resolution run LOW (left), the fiducial run MC1 (middle) and the high resolution run HR (right) at $t=18.4\,$kyr.}\label{fig_resolution}
\end{figure*}


We first consider the fiducial run MC1. Figure \ref{fig_turb100} shows three snapshots of the density, temperature, pressure and velocity from the run MC1. We adopt an isothermal MC background so the turbulent density causes the initial pressure fluctuations in the background, but the background pressure is much lower than the SNR pressure and can be ignored. The morphology of the SNR depends on the density distribution of the ambient MC. The forward shock front of the SNR shows some sub-pc finger structures due to the interaction between the SNR and the inhomogeneous MC. The SNR expands from  radii of $\sim 2-4$\,pc at $t=4.9\,$kyr to $\sim 4-6\,$pc at $t=24.6\,$kyr. The density in the interior region of the SNR decreases from $n_{\rm H} \sim 1\,$cm$^{-3}$ to $n_{\rm H}\sim 10^{-2}-10^{-1}$\,cm$^{-3}$, and the temperature and pressure drops from $\sim 10^7-10^8\,$K and $\sim (10^7 - 10^{8})\,k_{\rm B}\,$cm$^{-3}$\,K to $\sim 10^{6}-10^{7}\,$K and $\sim 10^{5}\,k_{\rm B}\,$cm$^{-3}$\,K because of radiative cooling. The interface zones between the SNR and the background MC has $T\sim 10^{4}\,$K. The velocities vary along different directions; in general, we find that the velocity deceases from $\sim 500$\,km\,s$^{-1}$ to $\lesssim 100\,$km s$^{-1}$ in three snapshots. These properties are comparable to those from the uniform case which will be discussed below. We still let the turbulent background evolve as the remnant expands. Since the SNR expansion timescale is 
\begin{eqnarray}
t_{\rm expand}\sim \frac{r_{\rm SNR}}{v_{\rm SNR}} &< & \frac{L}{2\,v_{\rm SNR}}\nonumber\\
&\sim& 7.8\times 10^{4}\,{\rm yr}\,\left(\frac{L}{16\,{\rm pc}}\right)\left(\frac{v_{\rm SNR}}{100\,{\rm km\,s^{-1}}}\right)^{-1},
\end{eqnarray} 
where $v_{\rm SNR}$ is the typical velocity of the outer SNR, and the definition of the average radius of the remnant $r_{\rm SNR}$ is introduced later in this section. The turbulent decay timescale is 
\begin{equation}
t_{\rm decay}\sim \eta t_s \sim4.8\times 10^{5}\,{\rm yr}\,\eta_{-2}\left(\frac{L}{16\,{\rm pc}}\right)\left(\frac{T}{30\,{\rm K}}\right)^{-1/2}.
\end{equation} 
where the typical value of $\eta$ is $\sim 10^{-2}-0.1$ (e.g., \citealt{Ostriker01}). Therefore we have $t_{\rm decay}\gg t_{\rm expand}$, and the background evolution does not affect the remnant evolution.

Figure \ref{fig_mixed} compares the morphologies of SNRs in various models at a fixed time 12.3\,kyr.  Compared to the fiducial, MC2 and MC3 runs in which the ambient medium has the same mean density, the adiabatic run shows the fastest expansion with significantly higher density and pressure in the interior region of the SNR. Radiative cooling leads to a smaller SNR,  by a factor less than two. In contrast to the runs with $\bar{n}_{\rm H}=100\,$cm$^{-3}$, run $\bar{n}10$ shows a larger SNR ($\sim 10\,$pc) with significant higher interior temperature ($\gtrsim 10^7\,$K) and faster expansion. For the same $\bar{n}_{\rm H}$, the morphology of the SNR in run MC2 is slightly different from that in run MC3 due to the different MC turbulent structure of the ambient MC. An initially smoother turbulent structure (lower $\mathcal{M}$) leads to a faster SNR expansion in some directions, with a smaller interface region ($T\sim 10^4$\,K). 


More quantitatively, we define some quantities to describe different gas components and measure the properties of SNRs. Following \cite{kim15}, we define the average radius of the SNR as calculated by $r_{\rm SNR} = \sum \limits_{\rm shell} \rho r (\Delta x)^{3}/\sum \limits_{\rm shell} \rho (\Delta x)^{3}$, where the shell region of the SNR is defined as the interface zones between the SNR and the MC background with $T <2\times 10^{4}\,$K but radial velocity $v_{r} > 1\,$km\,s$^{-1}$ for $\bar{n}_{\rm H}=100$ and 10\,cm$^{-3}$, and  $v_{r} > 10\,$km\,s$^{-1}$ for $\bar{n}_{\rm H}=1$\,cm$^{-3}$. The thermal and kinetic energies of a SNR are summed up over the zones with temperature higher than the background temperature, and the momentum feedback from a SNR to the ambient medium is measured by the radial momentum $p_{\rm SNR} =\sum \rho {\bf v}\cdot  {\bf r}(\Delta x)^{3}$. For the uniform medium case, \cite{kim15} using the \textsc{athena} code and \cite{walch15} using the SPH code \textsc{seren} compared their numerical results to the analytic results, and found good agreement. We also found similar results using the \textsc{athena++} code.  Figure \ref{fig_dyn1} shows the evolution of SNR radii $r_{\rm SNR}$, swept-up mass $M_{\rm sw}$, total (thermal and kinetic) energy and radial momentum of the SNR $p_{\rm SNR}$ in MC1, for the uniform and the adiabatic runs. The SNRs in MC1 and the uniform run show similar expansion behavior, as they have similar $r_{\rm SNR}$ and $M_{\rm sw}$. This result is different from that in \cite{martizzi15}, who found that a SNR in an inhomogeneous medium is larger than that in a uniform medium with the same mean density (their Fig. 3).  One of the main reasons for the difference is due to the different definitions of the remnant radii. In contrast to the averaged radius $r_{\rm SNR}$, Martizzi~et al. defined the outer shock radius of a remnant as the radius of the sphere enclosing 99\% of the total energy of the gas. Here, we define the ``maximum radius" of the shell region of a remnant with $v_r > 10\,{\rm km\,s^{-1}}$ ($r_{\rm max}$). The maximum radius of the shell region encloses almost the total energy of the remnant, and is equivalent to the radius given by Martizzi~et al . The left panel of Figure \ref{fig_Martizzi} shows the comparison of $r_{\rm SNR}$ and $r_{\rm max}$ for the models MC1, MC2, and MC3. We found that for all three models $r_{\rm max}\approx 10\,$pc at $t\sim 100\,$kyr, which is consistent with Martizzi~et al. However, the morphologies of the SNRs in our work are still different from those in Martizzi~et al., who adopted the lognormal density PDF from \cite{lemaster09b} with a power spectrum for spatial correlations to generate a turbulent density structure. In contrast to their method, we add velocity perturbations with a Gaussian profile into an initially uniform ambient medium, and generate density structure by velocity perturbations. In their models, Martizzi et al. found that the turbulent medium has low density channels, which allow faster SNR expansion in some preferred directions. The turbulent medium in our simulations does not show such preferred low-density channels, but the high-density filaments around a SNR slow down the SNR expansion almost in every direction. SNRs are more symmetric compared to those in Martizzi~et al. The result that $r_{\rm max}> r_{\rm SNR}$ is due to the large region of the shell around the remnants; $r_{\rm SNR}$ is better to measure the average size of remnants.


Figure \ref{fig_dyn1}  shows that the total energy in the SNR decreases faster in MC1, and the radial momentum is also lower compared to the uniform run. Because the SNR interacts with denser gas in MC1,  more energy is radiated away in the interface region. This result is consistent with \cite{martizzi15} (see their Fig. 4). On the other hand, the adiabatic run shows  significantly  larger $r_{\rm SNR}$ as an upper bound size of the SNR without cooling, largest swept-up mass and radial momentum $p_{\rm SNR}$. We cut off the adiabatic run at $\sim 50\,$kyr once most of the SNR interior energy escapes the computational box. The adiabatic run can be used to diagnose  our simulations: the turbulent structure is fractal in nature, so it does not add a new length scale, and the flow can be expected to approach an approximately self-similar flow with the ST solution  $r_{\rm SNR}\propto t^{0.4}$.  Figure \ref{fig_dyn1} shows that $r_{\rm SNR} \propto t^{0.4}$ for $t\lesssim 10\,$kyr; then $r_{\rm SNR}$ increases more slowly than the power-law of the ST solution. Deviation from the ST solution occurs because some SNR material starts to escape the computational box after 10\,kyr, so the averaged size of the SNR is smaller than the actual size.


Figure \ref{fig_dyn2} shows the remnant evolution for five runs: MC1, MC2, MC3, $\bar{n}_{10}$ and $\bar{n}_{1}$. We find that for the same $\bar{n}_{\rm H}$, different turbulent backgrounds give slightly different SNR evolution. The smoother the turbulent structure (lower $\mathcal{M}$) of the background, the slower radiative cooling occurs  in the SNR and the higher the momentum feedback $p_{\rm SNR}$. Interestingly, although the size of the SNR in MC2 is larger along some directions than that in run MC3, the averaged $r_{\rm SNR}$ and the swept-up mass are very similar in both runs. On the other hand, obvious differences can be observed for runs with various $\bar{n}_{\rm H}$, because lower $\bar{n}_{\rm H}$ leads to a larger SNR and faster expansion with slower radiative cooling. Note that the right panel of Figure \ref{fig_Martizzi} compares $r_{\rm max}$ to $r_{\rm SNR}$ -- we still use $r_{\rm SNR}$ to measure the size of remnants, but the values of $r_{\rm max}$ are only slightly larger than those of $r_{\rm SNR}$,  within a factor less than 1.4. Therefore we expect that the averaged size and shock velocity of a SNR can still be approximately described by the analytic estimate for the uniform case. Note that for the radiative cooling (snow plow) stage the radius of a SNR has the dependence $r_{\rm SNR}\propto E_0 ^{0.227} (\bar{n}_{\rm H})^{-0.263}$, where $E_0$ is the initial energy (\citealt{Draine11}, or see Appendix \ref{section_appendix} for details). Since the initial energy of a SNR cannot vary too much, but $\bar{n}_{\rm H}$ can change by several orders of magnitude, we conclude that $\bar{n}_{\rm H}$ is the controlling parameter for the size and expansion of a SNR. 


We also investigate the effects of spatial resolution. Figure \ref{fig_resolution} shows three density snapshots at the same time for the low-resolution run ($256^3$ zones), the fiducial case ($512^3$ zones), as well as the high-resolution run ($1024^3$ zones). It can be seen that the basic structure is similar in all three cases. More quantitatively,  we compare the evolution of the SNRs in the fiducial and the high-resolution runs, and find almost identical results. This is reassuring for the fidelity of our results for $512^3$ resolution. 
The numerical convergence of the remnant simulations have been discussed by \cite{kim15} and \cite{kim17a}, and we found similar results. For a uniform ambient medium, the possibility that radiative effects are overestimated and the simulations only slowly converge with increasing resolution to small scales cannot be  completely ruled out \citep[e.g.,][]{gentry18,steinberg18}. However, for a highly inhomogeneous background, the remnant simulations with spatial resolution of $\sim\,$pc or sub-pc cannot resolve the realistic field length \citep{Begelman90}, thus the cooling is dominated by the unresolved interface between the remnant and the ambient background. The turbulent mixing becomes more important than the unresolved microphysical mixing, and the numerical convergence is mainly due to the turbulent mixing at the interface (\citealt{kim17a}; Kim private communication). Nevertheless, even if the radiative cooling is overestimated, the adiabatic run shown in Figures \ref{fig_mixed} and \ref{fig_dyn2} gives the largest size of a SNR could be. The effect of radiative cooling is still less significant than the ambient mean density.

\begin{figure*}
\resizebox{\hsize}{!} {\includegraphics{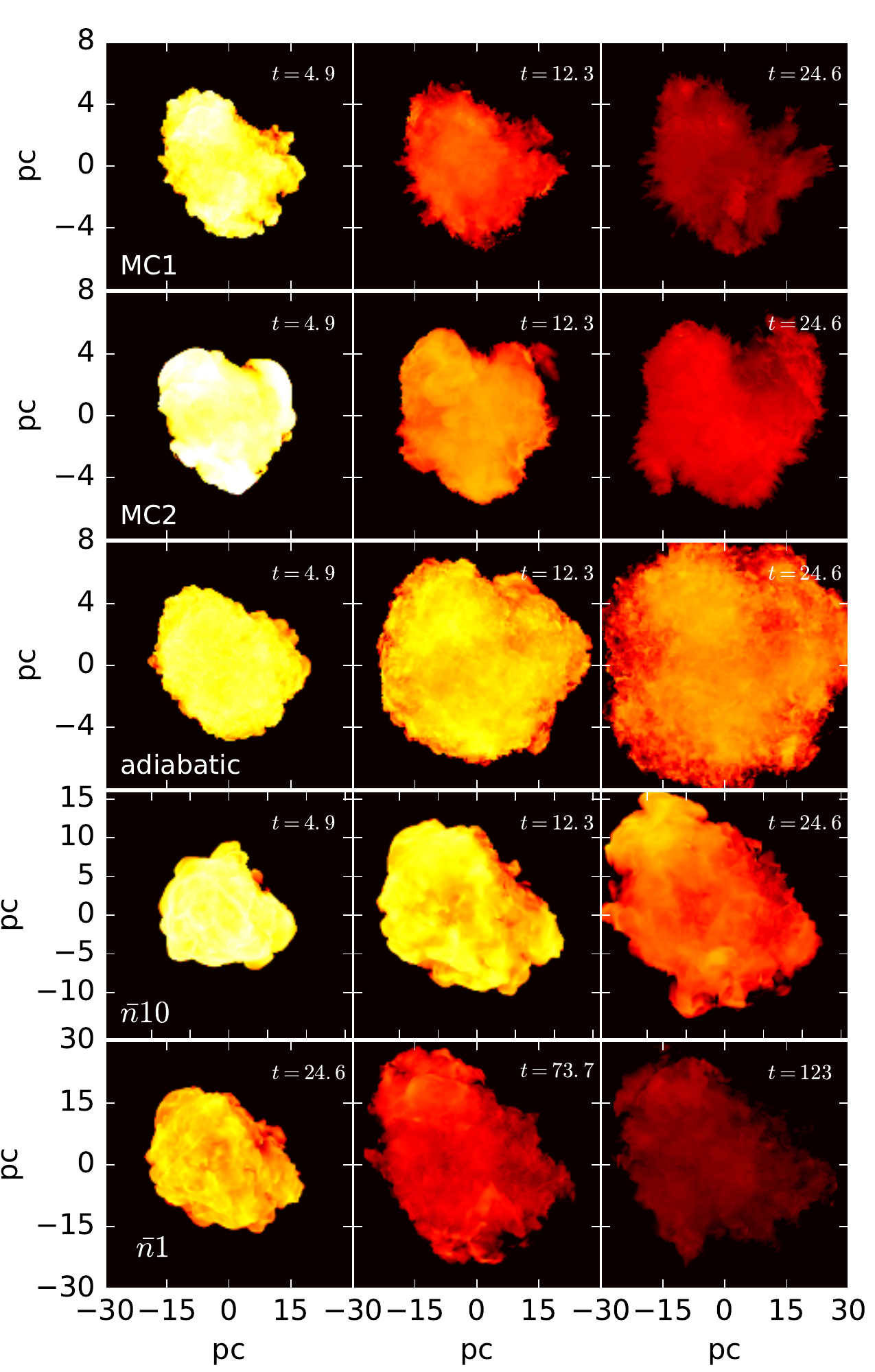}\includegraphics{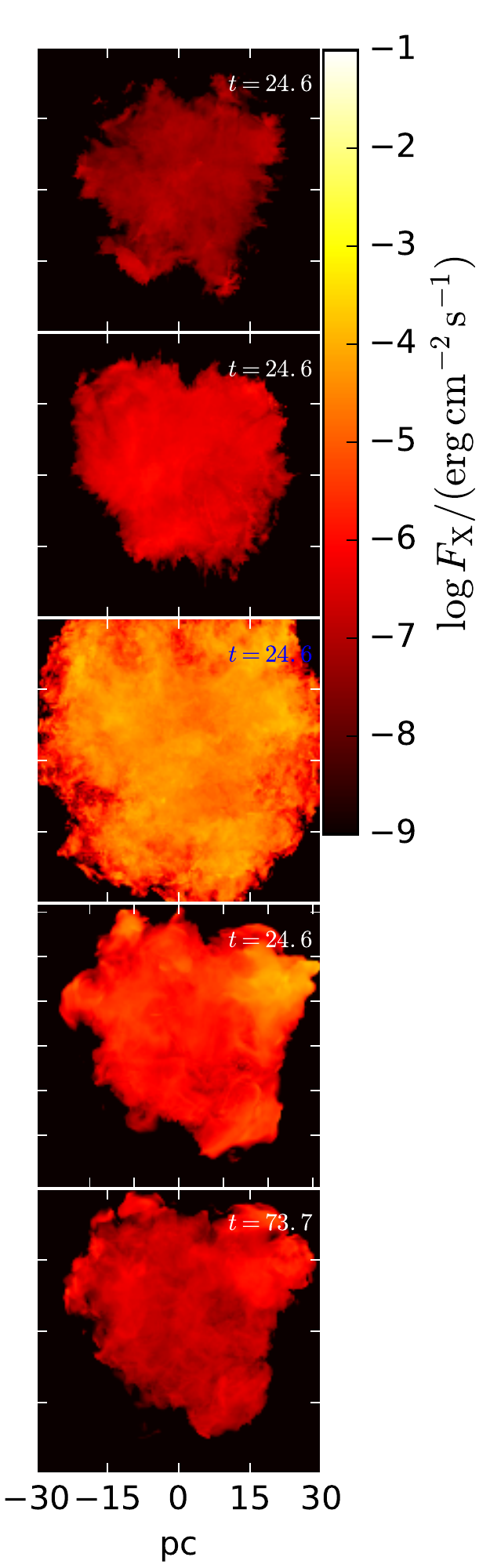}}
\caption{X ray flux 0.5-8 keV from SNR for runs MC1, MC2, adiabatic, $\bar{n}30$ and $\bar{n}1$. The first three columns are the flux along the $z$ direction for three different times, while the last column is the flux along the $x$ direction. Time is in units of kyr.}\label{fig_Xray}
\end{figure*}
\begin{figure*}
\centerline{\includegraphics[width=10cm]{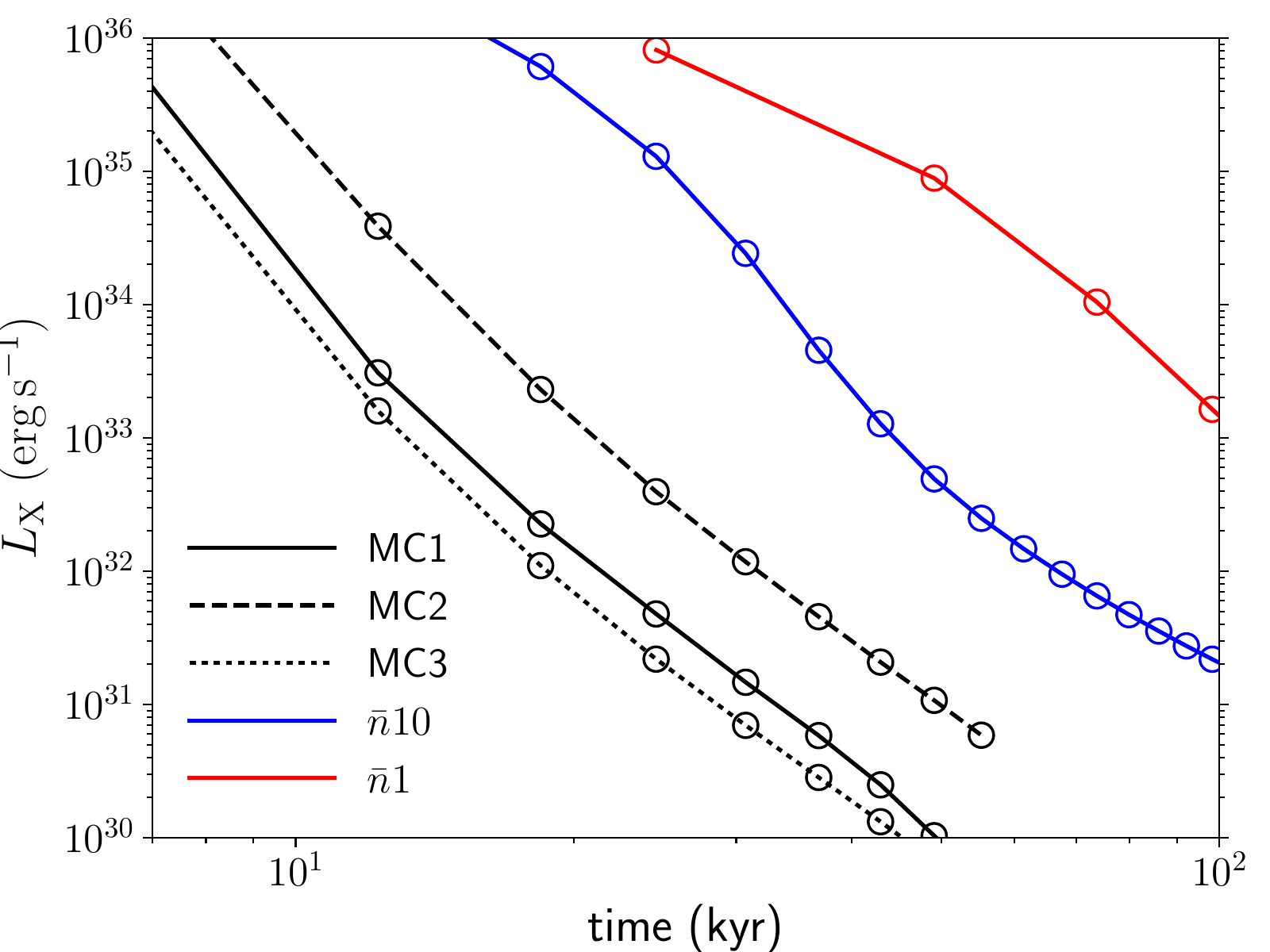}}
\caption{X-ray luminosity as a function of time for runs MC1, MC2, MC3, $\bar{n}10$ and $\bar{n}1$.}\label{fig_Xlum}
\end{figure*}

\section{X-ray Emission}\label{section_Xray}


X-ray observations are important in the study of SNRs. The shocked ambient medium can be heated  to produce thermal bremsstrahlung and line emission. Ignoring any non-thermal (synchrotron) component of X-ray emission, we can calculate the synthetic X-ray thermal fluxes from our simulations using the broadband emission introduced in Section \ref{section_cooling}. The X-ray continuum emission from each zone of the SNR region in the simulations is $e^{[\nu_1, \nu_2]} = n_{e}n_{\rm H}\Lambda(T)^{[\nu_1, \nu_2]}$, where $\Lambda(T)^{[\nu_1, \nu_2]}$ is the emissivity at temperature $T$ between the two (X-ray) frequencies $\nu_1$ and $\nu_2$ (see Figure \ref{fig_cooling}). Here we assume  solar metallicity, and $n_{e}$ is the electron density.  We calculate the X-ray fluxes along each direction of the computational box:
\begin{equation}
F_{X}^{[\nu_1, \nu_2]}\{x,y,z\}= \int n_{e}n_{\rm H}\Lambda(T)^{[\nu_1, \nu_2]} \{dx,dy,dz\},\label{lineofsight}
\end{equation}
where $F_{X}^{[\nu_1, \nu_2]}\{i\}$ is the integrated X-ray flux along the $i$ direction. We neglect the flux attenuation along the line of sight.

Figure \ref{fig_Xray} shows the images of the X-ray emission between 0.5\,keV to 8\,keV from the runs MC1, MC2,  adiabatic,  $\bar{n}10$ and $\bar{n}1$. The first three columns show the X-ray image snapshots along the $z$ direction at three times $t=4.9\,$kyr, 12.3\,kyr and 24.6\,kyr, except for the last run $\bar{n}1$, for which we show X-ray snapshots to 123\,kyr. The last column shows the X-ray snapshots along the $x$ direction at $t=12.3\,$kyr, except  the snapshot of run $\bar{n}1$ is at $t=73.7\,$kyr. Comparing X-ray images from runs MC1 and MC2, we find that for the same mean density $\bar{n}_{\rm H}$, the X-ray emission from the SNR in a less turbulent medium is brighter.  This is consistent with Figure \ref{fig_dyn2} which shows slower radiative cooling for a SNR in a less turbulent background. The size of X-ray images in each run is comparable to the 2D slices shown in Figures \ref{fig_turb100} and \ref{fig_mixed}. However, the finger structures of the interfaces in the 2D snapshots are mostly erased in the X-ray images due to the line of sight integration. In contrast to Figures \ref{fig_turb100} and \ref{fig_mixed} which show complex interface structure, there are clear and smooth boundaries between X-ray images and the ``X-ray-dark" background.


By integrating X-ray emission over the entire SNR evolution, we can obtain X-ray light curves. Figure \ref{fig_Xlum} shows X-ray luminosities $L_{X}$ for various models. The luminosities sensitively depend on the ages of the SNRs. Runs MC1, MC2 and MC3 show a very fast decreasing $L_{X}$, which drops 5 to 6 orders of magnitude in 50\,kyr. For runs $\bar{n}10$ and $\bar{n}1$, luminosities decrease slower than in the first three runs, but we still find $L_{X}$ dropping by almost 4 orders of magnitude in 70\,kyr for the least dense medium, $\bar{n}1$. Also, we find that the ambient turbulent structure plays an important role in shaping the X-ray luminosity. For the same  $\bar{n}_{\rm H}$, a smoother ambient background gives higher luminosity -- there is one order of magnitude difference between $L_{X}$ from MC2 and MC3. The X-ray light curves quantitatively analyze the images in Figure \ref{fig_Xray}, and we will use both the synthetic X-ray images and light curves to compare to observations.


\section{Comparison with Observations}\label{section_observations}

We compare our simulation results in Sections \ref{section_hydroresults} and \ref{section_Xray} to SNR observations, in particular to W44, which is interacting with a GMC within which it is embedded. Two other examples are W28 and IC 443, but they  appear to be only partially interacting with molecular gas. In addition to the observed size, age and X-ray emission of these SNRs, we also discuss radio observations of these SNRs. Radio combined with X-ray emission provide the morphological information on SNRs. More importantly, radio emission extends to the largest radius of a SNR and presumably delineates the forward shock front interacting with dense molecular gas.

\subsection{W44}

The SNR W44 is located in the Galactic Plane $(l,b)=(34.7,-0.4)$ at a distance about 3\,kpc from us (\citealt{Clark76,Wolszczan91,Taylor93}). The remnant is elongated, with radii of $11 \times 15$ pc with a mean radius of $\sim 13$ pc (\citealt{chevalier99}). W44 has long been observed to be interacting with dense molecular gas (e.g., \citealt{wootten77,Rho94}). The widespread OH masers \citep{claussen97} and other evidence of molecular interaction (see \citealt{reach05} and references therein) show that W44 is located in a well-defined GMC and propagating into the GMC, which has a radius of $\sim 58\,$pc and a total mass of $\sim 1.8\times 10^{6}\,M_{\odot}$, with a mean density of $\sim 60\,$cm$^{-3}$ (\citealt{Dame86}). Previous analytic work in the literature based on the scenario of SNR-clumps interaction estimated a much lower ambient interclump density to model W44. \cite{Rho94} found that W44 has an interclump ambient density of $n_{\rm H} \sim 0.09-0.26\,$cm$^{-3}$, which is far lower than the GMC mean density, and the SNR is in an adiabatic phase. Other work preferred that W44 is in the pressure-driven snow plow phase. \cite{Harrus97} fitted the interclump density to be $n_{\rm H} \sim 3-4 \,$cm$^{-3}$ with an initial energy of $\sim(0.7-0.9)\times 10^{51}\,$ergs. \cite{chevalier99} modeled W44 in a interclump medium with density $n_{\rm H} \sim 4-5 \,$cm$^{-3}$ with an initial energy of $10^{51}$ ergs, while \cite{cox99} suggested that the medium density is $\sim 6$\,cm$^{-3}$. All of these models give an order of magnitude lower ambient density than that of a typical GMC. The dense clumps may significantly increase the mean density of the overall surrounding medium around W44, but since the numbers and distribution of clumps are poorly understood, it is difficult to estimate the mean density of mixed clumps and an interclump medium. Also, none of these works considered turbulence of the GMC. 

The detection of X-rays from W44  was first reported by the \textit{Astronomical Netherlands Satellite} (\citealt{Gronenschild78}), then by many other X-ray satellites including \textit{Einstein}, the \textit{European X-Ray Observatory Satellite} (EXOSAT), the \textit{R\"{o}ntgen-Satellite} (ROSAT), the \textit{Advanced Satellite for Cosmology and Astrophysics} (ASCA), \textit{Chandra} and \textit{Suzaku} (e.g., \citealt{Szymkowiak80,Smith85,Jones93,Rho94,Harrus96,Harrus97,shelton99,shelton04,uchida12}). W44 is a typical SNR belonging to the class of mixed-morphology remnants. The X-ray images show that W44 is dominated by center-filled thermal X-rays, whereas the radio emission has an obvious shell  showing SNR-MC interaction (\citealt{claussen97,reach05}).  No discernible X-ray shell has yet been observed. \cite{Harrus96} showed that in the \textit{Einstein} band ($0.2-4\,$keV), the X-ray luminosity  is $4^{+30}_{-3}\times 10^{33}\,$ergs\,s$^{-1}$.

The age of W44 can be estimated as $\sim 20\,$kyr, which is obtained from the central pulsar in the SNR (\citealt{Wolszczan91}).  W44 is thus a middle-aged SNR. A comparison of the morphology of W44 with our simulations indicates that W44 is more like our $\bar{n}10$ model than other models.  
With an age of $\sim 20\,$kyr, the models with $\bar{n}_{\rm H}=100\,$cm$^{-3}$ give both $r_{\rm SNR}$ and $r_{\rm max}<10\,$pc, while the $\bar{n}10$ model gives $r_{\rm max}$ and $r_{\rm SNR} > 10\,$pc. The morphology of its X-ray image in Figure \ref{fig_Xray} shows an elongated structure with radii between $\sim 10-16\,$pc at $t=24.6\,$kyr, which is consistent with the morphology given by the radio image \citep[Fig.\ 2 of][]{claussen97}. If we consider that the total X-ray (0.5-8\,keV) is $\sim 10^{34}\,$ergs\,s$^{-1}$, the $\bar{n}10$ model in Figure \ref{fig_Xray} gives an age of $\sim 30\,$kyr, which is comparable to the age estimated for the center pulsar. Note that a more turbulent MC background leads to a lower X-ray brightness and younger age at a fixed $L_X$. This may indicate that W44 is embedded in a highly turbulent medium with $\bar{n}_{\rm H} \sim 10\,$cm$^{-3}$ and supersonic turbulent Mach number $\mathcal{M} \gtrsim 10$. 

A stronger SNR-MC interaction have been observed in the eastern part of the remnant (\citealt{Seta04}), which indicates the ambient medium has a higher mean density in this region compared to the western part. We expect that $\bar{n}_{\rm H} \sim 100^{-3}\,$cm$^{-3}$ can be reached near the eastern side of the remnant. \cite{Yoshiike13} observed that the atomic gas is $\lesssim 10\%$ of the molecular gas; thus the ambient medium W44 is molecular gas dominated, and our estimate $\bar{n}_{\rm H} \sim 10\,$cm$^{-3}$ is still valid for the most part of the ambient medium.

\begin{figure}
\centerline{\includegraphics[width=9cm]{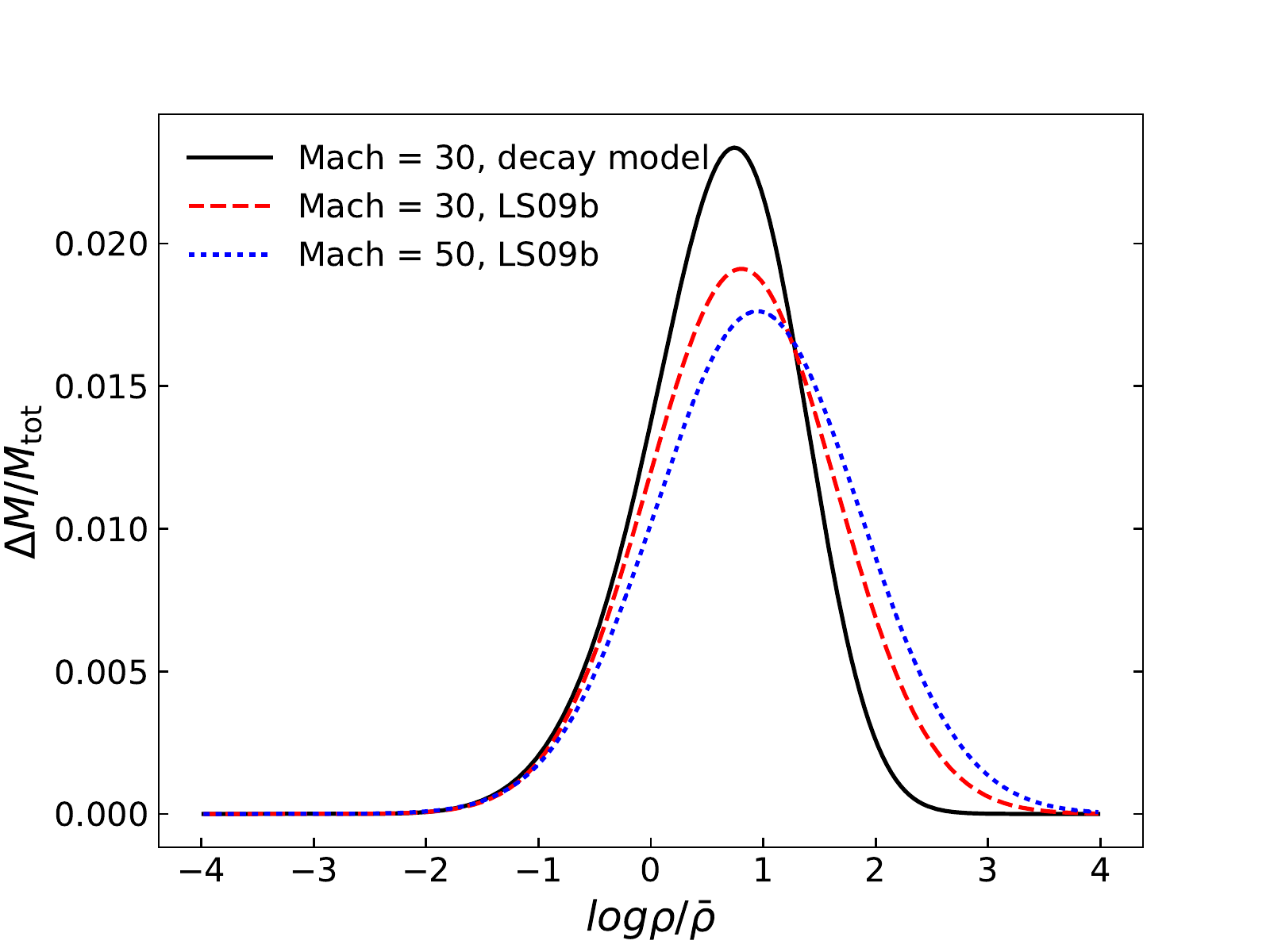}}
\caption{The density PDF of three turbulence model: the initial supersonic Mach number $\mathcal{M}=30$ (solid line) of the decay model, $\mathcal{M}=30$ (dashed line) and $\mathcal{M}=50$ (solid line) with the lognormal fit from \cite{lemaster09b} (LS09b).}\label{fig_Guassian}
\end{figure}

One may expect that the molecular gas with high density is in the filamentary structure of the turbulent medium with $\bar{n}_{\rm H}\sim\,10$cm$^{-3}$. Radio observations have traced the dense molecular gas interacting with the remnant with a density of $\sim 10^3-10^4\,$cm$^{-3}$, or even denser (\citealt{claussen97,reach05}). On the contrary, for $\bar{n}_{\rm H}=10\,$cm$^{-3}$, the most turbulent model MC3 ($\mathcal{M} =30$) in Section \ref{section_turb} only gives a small fraction of mass $\sim 0.7\%$ with density $\gtrsim 10^3\,$cm$^{-3}$. However, denser filaments can be generated from more turbulent media. Note that in Section \ref{section_turb} we adopted the decaying model to generate the turbulent MC. It is possible that some extra sources maintain the turbulent energy in the MC, so the turbulent structure follows an ``energy injection model" as described by \cite{lemaster09a,lemaster09b}, in which energy is provided into the medium in each simulation timestep. Figure \ref{fig_Guassian} compares the density PDF from our decaying model MC3 to the energy injection model with $\mathcal{M}=30$ and 50. The lognormal density PDF for the ``energy injection model" has a longer tail for high density gas than the decaying model. Neither the decaying model nor the energy injection model generated turbulence with low-density channels along some preferred directions, so we expect a SNR evolving in a highly turbulent medium generated by the energy injection model still has a comparable size as that in a uniform medium with the same $\bar{n}_{\rm H}$.  Another possibility to create very dense molecular filaments with $\bar{n}_{\rm H}=10\,$cm$^{-3}$ is due to the self-gravity of the MC, which leads to the collapse of the turbulent dense filaments and creates denser filaments and molecular cores compared to those do not include self-gravity. 

In short, both adding energy injection into a highly turbulent medium or adding self-gravity may generate molecular filaments with high density $\sim 10^3-10^4\,$cm$^{-3}$ with mean density $\bar{n}_{\rm H}=10\,$cm$^{-3}$. This estimated mean density is slightly higher than the estimated interclump density by previous analytic work, but note that our turbulence model is different from the analytic clump model for the MC. The mean density of the surrounding medium based on the clump model may increase by increasing the number of clumps, but our turbulence model fixed the mean density of the surrounding medium, which is about six times lower than the GMC mean density $\sim 60\,$cm$^{-3}$. A question then arises: if the MC turbulent structure is correct, why is the density of the ambient medium much lower than the mean density of the host GMC? One possible explanation is that feedback from W44's progenitor including radiation and stellar wind has already affected the ambient molecular gas and decreased its mean density. We expect our predictions of the MC properties around W44 can be tested by new observations. For example, future observations using the Atacama Large Millimeter Array ({\it ALMA}) can probably reveal highly resolved structure of the MC on the western side of W44. 



Moreover, \cite{cox99} discussed the impact of thermal conduction and estimated that the central density and temperature of W44 are $\sim 0.7\,$cm$^{-3}$ and $\sim 7\times 10^6\,$K respectively. In our simulations we neglect thermal conduction, so the central density is one order of magnitude lower but the temperature one order of magnitude higher than in \cite{cox99}. In order to directly compare the synthetic X-ray emission to highly resolved mixed morphology observations and explore the origin of mixed morphology,  further work needs to be done to include thermal conduction in SNR simulations (see  discussion  in Section \ref{section_conclusions}).

\subsection{W28}

The SNR W28 (or G6.4$-$0.1) is located at a distance between 1.6 to 4\,kpc in the Galactic Plane $(l,b)=(6.71,-0.05)$ (\citealt{Goudis76,Lozinskaya81,Velazquez02}). The angular diameter of W28 is $\sim 50'$, corresponding to a radius of $\sim 15\,$pc$\,(D/2\,$kpc).  The size of W28 is comparable to that of W44. The estimated age of W28 spans from 33\,kyr to 150\,kyr (\citealt{Kaspi93,Velazquez02,Rho02,Zhou14}), but it is most likely that its age is between $30-45\,$kyr (\citealt{Giuliani10,Zhou14}). The radio structure of W28 has been investigated for four decades (\citealt{Kundun70,Shaver70,Milne71,Goudis76}), and it has been observed to show  MC interaction through OH masers lines, CO lines and other molecular lines implying shocked gas  (e.g.,\citealt{wootten81,frail94,frail98,claussen97,claussen99,Arikawa99,Dubner00}). The molecular interaction can only be  observed in the northeastern part of the remnant, and the parent GMC with which W28 is interacting has a diameter of $\sim 25\,$pc and a mass of $\sim 1.4\times 10^{6}\,M_{\odot}$, corresponding to a mean density of $\sim 600\,$cm$^{-3}$ (\citealt{Dame01,reach05}).

X-ray observations of W28 have been carried out by several generations of X-ray satellites including \textit{Einstein}, ROSAT, ASCA, XMM-Newton, \textit{Chandra} and \textit{Suzaku} (\citealt{Long91,Rho02,Keohane05,Kawasaki05,sawada12,pannuti17}). The X-ray maps show a primarily center-filled thermal component with partial shells at the northeast and southwest of the SNR. Therefore W28 also has a mixed morphology. The estimated total X-ray luminosity by ROSAT is $\sim 6\times 10^{34}\,$ergs\,s$^{-1}$ (\citealt{Rho02}), much brighter than W44. 

We compare both the X-ray data and the size of W28 to our simulations. Since the age is uncertain, we can compare W28 to all the models in Figure \ref{fig_Xlum}. For $L_X = 6\times 10^{34}\,$ergs\,s$^{-1}$, runs with $\bar{n}_{\rm H}=100\,$cm$^{-3}$ give an age $\lesssim 10\,$kyr, which is too young to match the observation. Also for $\bar{n}_{\rm H}=100\,$cm$^{-3}$ and an age of $30-45\,$kyr, the average size of a SNR is $\sim 5-6\,$pc with a maximum radius $r_{\rm max}\sim 10\,$pc, which are all smaller than the observed remnant. The models with $\bar{n}_{\rm H}=100\,$cm$^{-3}$ are ruled out. Run $\bar{n}10$ in Figure \ref{fig_Xlum} constrains the age of SNR to $\simeq 26\,$kyr, which is still slightly younger than W28, while run $\bar{n}1$ gives an age of $\simeq 55\,$kyr, which is somewhat older than W28's age. However, according to Figures \ref{fig_dyn2} and \ref{fig_Xray}, the $\bar{n}1$ model with an age of $\simeq 55\,$kyr has a much larger size than the radius of W28 ($\sim15\,$pc). The $\bar{n}1$ model is unlikely to describe W28. On the other hand, we expect that a SNR in a medium with a mean density of $\sim 10\,$cm$^{-3}$ and a turbulent structure similar to MC2 to be consistent with the combination of age and X-ray data of W28. Note that run $\bar{n}10$ in Figure \ref{fig_Xray} shows an elongated structure, in particular, a longest radius of $\sim 15\,$pc at $t=24.3\,$kyr. A less turbulent medium gives a more spherical morphology but  brighter X-ray emission.  We expect that replacing run $\bar{n}10$ with a less turbulent background would show a more spherical morphology and a radius of $\sim 15\,$pc at at $t\sim 30$kyr. 

One may find an issue that the above estimate assuming that the remnant is embedded in a GMC. Since W28 is only partially interacting with its host GMC, the mean density we estimate may not be for the MC, but for the atomic medium which around the remnant. For the atomic medium, turbulence is probably not important. However, note that the distance between the remnant center to the northeastern shell shown by X-ray observation (\citealt{Rho02}, see their Figs. 1-5) spans $\sim 20'-25'$, which is comparable to the averaged radius of the remnant if the SN exploded near the center of the remnant. We expect that the density of the MC is also comparable to that of the ambient atomic medium.  Since the SNR-MC interaction region is brighter with a shell X-ray emission, we expect that the MC has low turbulence since a smoother MC gives brighter X-ray emission. This result may conflict with radio observations which also traced dense molecular gas with $\sim 10^3-10^4\,$cm$^{-3}$ as W44. In this case, the high-density molecular gas cannot be modeled by the filaments of the MC turbulence. Clumpy molecular clouds with a low density interclump molecular medium are probably more realistic to explain the observations. 


To summarize, the observations of W28 can be modeled as a SNR partially interacting with a MC with a density of $\bar{n}_{\rm H}\sim 10$\,cm$^{-3}$ and a supersonic turbulent Mach number $\mathcal{M} \lesssim 3$, if the SN occurred near the center of the remnant. Since the average density of the host GMC is large ($\sim 600\,$cm$^{-3}$), the question why the estimated mean density around W28 is much lower than that of its parent GMC is even more serious than that for W44. In addition to the scenario of strong stellar feedback, another possibility is that the SN actually occurred near the interface between the turbulent MC gas and lower density atomic gas. We have found that a 10$^{51}$\,ergs SN in  a typical MC with $\sim 100\,$cm$^{-3}$ expands only to a radius of $5-6$ pc in 30\,kyr, so the molecular clumps with a density of $10^3-10^4\,$cm$^{-3}$ can be explained as the filaments of the turbulent MC. The other part of the SN energy then drives a larger shocked region in an HI envelope that may be surrounding the MC.  The outer contours of radio emission then delineate the forward shock front of the SNR.  \cite{iffrig15} have made a start of modeling a SN explosion close to the edge of a turbulent MC.  More detailed simulations are needed that can be compared to observed remnants.

\subsection{IC 443}

IC 443 (G189.1+3.0) has a distance of 1.5\,kpc with a diameter of 45$'$ (\citealt{Lozinskaya81,Fesen84}), corresponding to a radius of $\sim 10\,$pc. IC 443 is another example of SNR interaction with molecular gas. The MC in front of the SNR was first discovered by \cite{Cornett77}, and the SNR-MC interaction and shocked molecular gas have been discovered and confirmed by OH masers, CO, H$_{2}$ and other molecular features (e.g., \citealt{DeNoyer78,DeNoyer79a,DeNoyer79b,Burton88,Ziurys89,Dickman92,claussen97,Rho01,Hewitt06,Xu11,Lee12}). The remnant shows a complex ambient environment. Only the central part of the remnant is propagating into molecular gas, while the northeastern part is interacting with atomic clouds (\citealt{Lee08,Lee12}). 

A variety of X-ray observations have been carried out by \textit{Einstein}, \textit{Ginga}, ROSAT, ASCA, XMM-Newton, \textit{Chandra} and {\it NuSTAR} to show that IC 443 has an extended thermal X-ray emission with some point sources (\citealt{Petre88,Wang92,Asaoka94,Kawasaki02,Troja06,Zhang18}). The early X-ray observations suggested that the remnant is very young (\citealt{Wang92,Petre88}), but measurements of the transverse velocity of a central pulsar show that the remnant's age is $\sim 30\,$kyr (\citealt{Olbert01}). IC 443 is also categorized as a mixed-morphology remnant, since the X-rays show a center-filled structure. The X-ray luminosity from 0.2 to 4\,keV is estimated to be $\sim 5 \times 10^{33}$\,ergs\,s$^{-1}$ (\citealt{Olbert01}).

The SNR-MC interaction region has a diameter of 20$'$, corresponding to a radius of 4.4\,pc, which is much smaller than in W44 and W28. We estimate that the mean density of the MC is about $\sim 100\,$cm$^{-3}$, which is the typical MC density. It is uncertain how much X-ray emission is contributed by the SNR-MC interaction. A simple estimate gives the emission as $L_{X}\lesssim 5\times 10^{33}(20'/45')^{2}\lesssim 10^{33}\,$ergs\,s$^{-1}$. Note that the most X-ray bright region is in the atomic region, the real X-ray luminosity from the MC region may be lower than our estimate. However, Figure \ref{fig_Xlum} shows that the fiducial and MC3 run give $L_{X}\sim$a few\,$\times 10^{31}\,$ergs\,s$^{-1}$ at $t=30\,$kyr, which is much lower than the observed X-rays. The MC2 run shows $\sim 10^{32}\,$ergs\,s$^{-1}$, thus we suggest that the MC2 run is closer to the real environment in front of IC 443.  


Also, we estimate that for the atomic medium a mean density of $\sim 30$\,cm$^{-3}$ with the same turbulent structure as MC3 is consistent with both the age and the X-ray luminosity of IC 443.  This estimate is basically consistent with \cite{chevalier99}, who estimated the density of ambient medium as $\sim 15\,$cm$^{-3}$. The remnant is likely to occur near a MC, expanding into the high-density filaments of the MC, then propagating into the low-density surrounding atomic medium.


\section{Conclusions and Discussion}\label{section_conclusions}

In this paper we investigate whether the interaction between core-collapse SN explosions and turbulent MCs show observational evidence that can reveal the properties of the  surrounding medium. We perform a series of 3D hydrodynamic simulations to model the evolution of individual SNRs with radiative cooling in an inhomogeneous medium and supersonic turbulence. The turbulent structure is generated by initial velocity perturbations with a Gaussian random distribution and a Fourier power spectrum $|v^{2}(k)|\propto  k^{-4}$.  The density and volume PDFs of the medium then follow  lognormal distributions, as shown in equation (\ref{lognormal}). The properties of the turbulent medium are determined by the initial turbulent Mach number $\mathcal{M}$. We generate three turbulence models with $\mathcal{M}=10$ (MC1), $\mathcal{M}=3$ (MC2) and $\mathcal{M}=30$ (MC3). 

Following \cite{kim15}, the expansion of a SNR in an inhomogeneous medium can be measured by the average radius $r_{\rm SNR} = \sum \limits_{\rm shell} \rho r (\Delta x)^{3}/\sum \limits_{\rm shell} \rho (\Delta x)^{3}$, where the SNR shell region is defined as the zones with temperature $T <2\times 10^{4}\,$K but radial velocity $v_{r} > 1\,$km\,s$^{-1}$ for the mean density of the surrounding medium $\bar{n}_{\rm H}=100$ and 10\,cm$^{-3}$, and $v_{r} > 10\,$km\,s$^{-1}$ for $\bar{n}_{\rm H}=1\,$cm$^{-3}$. We find that $r_{\rm SNR}$ is mainly controlled by $\bar{n}_{\rm H}$ (Fig. \ref{fig_dyn2}). For a fixed $\bar{n}_{\rm H}$, the average radius $r_{\rm SNR}$ only slightly changes due to the different turbulent structure of the surrounding medium (Fig. \ref{fig_dyn1}). Therefore, the analytic estimate of $r_{\rm SNR}$ for a uniform surrounding medium can  approximately describe the size of SNR with $r_{\rm SNR}\propto E_0^{0.227}(\bar{n}_{\rm H})^{-0.263}$, where $E_0$ is the initial energy of the supernova. This conclusion is different from that in \cite{martizzi15}, who found  faster expansion of a SNR in a more turbulent medium. The main reasons of the difference are caused by different definitions of the remnant sizes and the turbulent structures. We find that the maximum radius of the remnant shell region is comparable to those in Martizzi~et al. with the same $\bar{n}_{\rm H}$. However, Martizzi~et al. used the lognormal density PDF with a power spectrum for spatial correlations to generate the turbulent density, while we use an initial Gaussian velocity perturbation in a uniform medium to observe the evolution of turbulence. The low-density channels along some preferred directions in Martizzi~et al. help a SNR propagate faster in these channels than other high-density region, but we find that in our turbulence model there are no such channels along special directions.  The high-density filaments around a SNR eventually slow down the SNR in almost each direction. Our turbulence model is also different from both the multiphase ISM model (\citealt{kim15,kim17a,kim17b}), and the clumpy cloud models discussed by some analytic and numerical work (\citealt{white91,chevalier99,cox99,Ferrand14}).  

For fixed $\bar{n}_{\rm H}$, we find that a SNR in a more turbulent medium has stronger radiative cooling and less momentum feedback into the surrounding medium compared to that in a less turbulent medium. These results are consistent with the synthetic X-ray emission from SNRs. Using the CIE broadband X-ray emission for simplicity, we simulate the synthetic X-ray images (Fig. \ref{fig_Xray}) and light curves $L_X$ (Fig. \ref{fig_Xlum}) from our simulations. Although the X-ray luminosity still mainly depends on $\bar{n}_{\rm H}$, the turbulent structure of the surrounding medium may cause $L_X$ to vary by an order of magnitude  between a SNR in a highly turbulent medium compared to that in a uniform medium. A more turbulent medium leads to  dimmer X-ray emission from the SNR. Moreover, in contrast to the 2D slices with complex finger structure in the interface, the X-ray images show clear and smooth boundaries between SNRs and the ambient MC background. This is because the line-of-sight integration (equation \ref{lineofsight})  smooths out the finger structure shown by 2D slices.

We compare our simulations to observed SNRs, in particular, to W44, W28 and IC 443. W44 is embedded in a GMC and propagating into it, making it a good candidate to be compared to the simulations. Since W28 and IC 443 appear to be only partially interacting with their host GMCs, we  give only  rough constraints on their MC environment. Since W44 is an elongated remnant with radius of $11\times 15\,$pc and an age of $\sim 20\,$kyr, we estimate that the MC around W44 has a mean density of $\sim 10\,$cm$^{-3}$, which is lower than the mean density of the global GMC. The low X-ray luminosity of W44 indicates that W44 may be propagating in a highly turbulent medium with turbulent Mach number $\mathcal{M}\gtrsim 10$.  The molecular gas with high density $\sim 10^3-10^4\,$cm$^{-3}$ can be explained as the filaments of the turbulent MC. For W28, we estimate that its ambient medium has $\bar{n}_{\rm H}\sim 10\,$cm$^{-3}$, but is less turbulent ($\mathcal{M}\lesssim 3$) compared to the medium around W44 ($\mathcal{M} \gtrsim 10$).  Since part of W28 is out of the MC region, the estimated ambient density may refer to the atomic medium. If the SN occurred at the center of the remnant, we expect that the MC has a similar mean density as the whole surrounding medium. Another possibility is that the SN occurred close to a MC with a typical density of $\sim 100\,$cm$^{-3}$ and expanding into a low-density atomic medium. This is the ``shock breakout" scenario for the remnant. 
For IC 443, we estimate that the MC in front of the remnant has $\bar{n}_{\rm H}\sim 100\,$cm$^{-3}$, which is the typical density of MC. 
The density of the ambient medium around W44 and W28 is significantly lower than that of global GMCs. The lower density may indicate that stellar feedback including radiation and stellar winds from the SNRs' progenitors may play an important role in shaping the environment of massive stars. We expect future high-resolution observations using {\it ALMA} to better probe the properties of the GMCs associated with these SNRS, and to test our turbulent models of MC interaction.


There are many directions that are beyond the scope of our paper and worth exploring. One direction is the addition of thermal conduction into SNR simulations. Since we do not directly attempt to use the synthetic X-ray emission from our simulations to explain the phenomena of mixed morphology, we  neglect thermal conduction in our work. A relevant work has been done by \cite{Ferrand14}, who simulate the interaction between SNRs and clumpy clouds which are evaporated by thermal conduction, and test the analytic model of \cite{white91}. Although the evaporation of clumpy clouds  probably does not significantly contribute to the center-filled X-ray emission, thermal conduction in the hot interior SNR may be important. \cite{Harrus97} found that conduction is necessary to fit X-ray emission from W44, and \cite{cox99} proposed that conduction may dominate energy transport in W44 and prevent formation of a low density cavity inside the remnant. The center-filled morphology of X-ray emission from W44 may be explained by thermal conduction in the remnant. However, no numerical simulations have been carried out to test Cox et al.'s scenario and explore the impact of conduction.   



Another issue is the treatment of broadband X-ray emission. So far we use the CIE broadband emission as an approximation to calculate X-ray emission, but a more realistic model of X-ray emission from SNR-MC interaction should use NEI instead of CIE emission especially in the interface region with $T<10^{6}\,$K. We encourage future simulations to generate synthetic X-ray emission using NEI emission and directly compare to the high-resolution X-ray observations. \cite{Orlando16} used the NEI emission model VPSHOCK available in the XSPEC package along with the NEI version 2.0 atomic data from ATOMDB (\citealt{Smith01}) to synthesize the X-ray emission from SN 1987A.

Our turbulence models do not include self-gravity and magnetic fields. The timescale of gravitational contraction is $t_{\rm cc}\sim (\pi/G\bar{\rho})^{1/2}\sim 5.3\,$Myr\,$(\bar{n}_{\rm H}/10^3\,{\rm cm}^{-3})$, which is comparable to the lifetime of the SNR progenitor stars. Therefore self-gravity can be important to model the collapse of the high-density clumps or filaments in the turbulent medium. Numerical simulations have revealed that as self-gravity is included, the gas density of a turbulent medium still follows a lognormal distribution, with the addition of a high-density power-law tail (\citealt{Kritsuk11,Pan16}). Self-gravity is also worthwhile to be further discussed in the context of the formation of high-density molecular filaments. Also, the lognormal distribution of magnetohydrodynamic turbulence may be different from that without magnetic fields (\citealt{Li04,Li08,lemaster09b}). Magnetic fields may not be significant for the evolution of a middle-aged SNR (\citealt{kim15}), but the interior magnetic field structure inside the remnant is crucial to modeling radio emission. Cosmic rays can increase the pressure of the remnants and boost the late time SNR expansion and momentum injection \citep{Diesing18}. Future studies also need to consider magnetic fields, cosmic rays,  and comparisons to radio observations.

\section*{Acknowledgments}

We thank the referee, Davide Martizzi, for thoughtful comments for us to improve the paper. DZ thanks C.~Y.~Chen and E.~Ostriker for sharing the \textsc{athena} code for the turbulence driver, C.~G.~Kim for comments on the text, and S.~Davis, M. Ruszkowski, and S. Oey for helpful discussion. This work used the Extreme Science and Engineering Discovery Environment (XSEDE), which is supported by National Science Foundation (NSF) grant No. ACI-1053575. We also used the computational resources provided by the Advanced Research
Computing Services (ARCS) at the University of Virginia, and computational sources from S. Davis. DZ acknowledges support from NSF grant AST-1616171, and RAC 
from the National Aeronautics and Space Administration through Chandra Award Number TM6-17004X issued by the Chandra X-ray Center, which is operated by the Smithsonian Astrophysical Observatory for and on behalf of the National Aeronautics Space Administration under contract NAS8-03060.

\begin{appendix}

\section{Analytic Estimate of a SNR Evolution with Radiative Cooling}\label{section_appendix}

We consider a SNR expands in a uniform medium. The initial energy of the SNR is $E_0$ and the density of the ambient medium is $n_{\rm H}$. After the SNR moving into the ST stage, the entire SNR can be described by the ST self-similar solution. The shock radius, velocity and temperature behind the postshock are given by (\citealt{Draine11}, see also \citealt{kim15})
\begin{eqnarray}\label{eq:ST}
&&r_{\rm SNR} = 3.15\,{\rm pc}\;E_{0,51}^{1/5}  n_{\rm H,1}^{-1/5} t_3^{2/5},\\
&&v_{\rm SNR} = 1.23\times 10^3\,{\rm km\;s^{-1}}\; E_{0,51}^{1/5} n_{\rm H,1}^{-1/5} t_3^{-3/5},\\
&&T_{\rm SNR} = 2.09\times10^7\,{\rm K}\; E_{0,51}^{2/5} n_{\rm H,1}^{-2/5} t_3^{-6/5} ,
\end{eqnarray}
where $E_{0,51}=E_0/10^{51}\,$ergs, $n_{\rm H,1}=n_{\rm H}/10$\,cm$^{-3}$, and $t_3=t/10^3\,$yr. According to \cite{Draine11}, if the radiative cooling function in a range of $10^5\,{\rm K} < T < 10^{7.5}\,{\rm K}$ is simply treated with a power-law formula
\begin{equation}
\Lambda \approx C(T/10^{6}\,{\rm K})^{-0.7}n_{\rm H}n_{e}
\end{equation}
with $n_{e}$ being the electron number density in the ambient medium and $C=1.1\times 10^{-22}\,$ergs\,cm$^{-3}$\,s$^{-1}$, the cooling time, shock radius, velocity, and the temperature behind the postshock when the SNR transfers from the ST stage to the snow plow stage are given by
\begin{eqnarray}\label{eq:cool}
&&t_{\rm rad} = 12.4\,{\rm kyr}\; E_{0,51}^{0.22} n_{\rm H,1}^{-0.55},\\
&&r_{\rm rad} = 8.59\,{\rm pc}\; E_{0,51}^{0.29} n_{\rm H,1}^{-0.42}, \\
&&v_{\rm rad}  = 272\,{\rm km\;s^{-1}}\; E_{0,51}^{0.07} n_{\rm H,1}^{0.13},\\
&&T_{\rm rad}  = 1.03\times10^6\,{\rm K}\; E_{0,51}^{0.13} n_{\rm H,1}^{0.26}.
\end{eqnarray}
Note that for $n_{\rm H}=100\,$cm$^{-3}$, we have $t_{\rm rad}\approx 3.5\,$kyr\,$E_{0,51}^{0.22}$, which is consistent with the uniform run (Fig. \ref{fig_dyn1}) in the paper. 

During the snow plow stage driven by thermal pressure of the SNR, the gas in the SNR hot interior can be treated approximately as in an adiabatic process and the pressure of the SNR has $P_{\rm SNR}v_{\rm SNR}^{\gamma}$, or $P_{\rm SNR}\propto r_{\rm SNR}^{-3\gamma} =r_{\rm SNR}^{-5}$. The pressure of the SNR shell satisfies
\begin{eqnarray}
\frac{d}{dt}(M_{\rm sw}v_{\rm SNR})&\approx& P_{\rm SNR}4\pi r_{\rm SNR}^{2}\nonumber\\
 &= & 4\pi P_{\rm SNR}(t_{\rm rad})r_{\rm rad}^{5}r_{\rm SNR}^{-3}\label{eq:momentum}
 \end{eqnarray}
Assuming the evolution of $r_{\rm SNR}$ still obeys a power low $r_{\rm SNR} \propto t^{\eta}$, we use equation (\ref{eq:momentum}) to solve $\eta$ and find $\eta=2/7$. Therefore
\begin{eqnarray}
r_{\rm SNR} \approx r_{\rm rad}\left(\frac{t}{t_{\rm rad}}\right)^{2/7}\approx 8.08\,{\rm pc}\;E_{0,51}^{0.227}n_{\rm H,1}^{-0.263} t_{4}^{2/7},\label{eq:radius}
\end{eqnarray}
where $t_4 =t/10^{4}\,$yr, and 
\begin{eqnarray}
v_{\rm SNR}&\approx& \frac{2}{7}\frac{r_{\rm SNR}}{t}\approx \frac{2}{7}\frac{r_{\rm rad}}{t_{\rm rad}}\left(\frac{t}{t_{\rm rad}}\right)^{-5/7}\nonumber\\
&\approx& 226\,{\rm km\,s^{-1}}\;E_{0,51}^{0.227}n_{\rm H,1}^{-0.263}t_{4}^{-5/7}\label{eq:vel3}.
\end{eqnarray}
We use equation (\ref{eq:radius}) in Section \ref{section_hydroresults} to estimate the size of a SNR both in uniform and turbulent medium. 

We do not explore the SNR evolution at later time in this paper. In order to complete the analytic estimate we also discuss the last stage which is the momentum-driven snow plow stage. This stages stops once the shock velocity is comparable to the sound speed of the medium. Then the SNR fades away in the ambient medium. Setting $v_{\rm SNR}$ in equation (\ref{eq:vel3}) to be the sound speed $c_s$, one derives the time and radius of the SNR when it fades away as
\begin{eqnarray}
&&t_{\rm fade} \approx 0.79\,{\rm Myr}\;E_{0,51}^{0.318}n_{\rm H,1}^{-0.368}(c_s/10\,{\rm km\;s^{-1}})^{-7/5},\\
&&r_{\rm fade} \approx 28.1\,{\rm pc}\;E_{0,51}^{0.318}n_{\rm H,1}^{-0.368}(c_s/10\,{\rm km\;s^{-1}})^{-2/5}.
\end{eqnarray}

\end{appendix}

\end{document}